\documentclass[12pt]{article}

\RequirePackage[T1]{fontenc}
\usepackage[utf8]{inputenc}

\usepackage[height=8.85in,width=6.45in]{geometry}
\renewcommand{\baselinestretch}{1.3}

\usepackage{titlesec}

\titleformat*{\section}{\large\bfseries}
\titleformat*{\subsection}{\bfseries}

\usepackage[svgnames]{xcolor}
\definecolor{refcolor}{rgb}{0.3,0.3,1}

\usepackage[
  bookmarks=false,
  colorlinks=true,
  citecolor=refcolor,
  linkcolor=refcolor,
  urlcolor=magenta
]{hyperref}

\usepackage[hang]{caption}
\usepackage{subcaption}

\usepackage[hang,bottom]{footmisc}
\usepackage{appendix}

\usepackage{amsmath}
\numberwithin{equation}{section}

\usepackage{amssymb}
\usepackage{bm} 
\usepackage{bbm} 
\usepackage{braket}
\usepackage{cancel}
\usepackage{mathdots} 
\usepackage{mathtools} 
\usepackage{nicematrix}
\usepackage{extarrows}

\usepackage{graphicx}
\usepackage{pifont} 
\usepackage{marvosym} 

\usepackage{ascmac} 
\usepackage{fancybox} 

\usepackage{array}
\usepackage{longtable}
\usepackage{arydshln}
\usepackage{multirow}

\usepackage{tikz}
\usetikzlibrary{quantikz2} 
\usetikzlibrary{shapes.geometric}
\usetikzlibrary{matrix}
\usetikzlibrary{arrows.meta}

\newcommand{\hexlctrl}[2]{%
  \gate[style={shape=circle, 
               draw, fill=white, inner sep=-2pt}][6pt][6pt]{#2}%
  \ifnumcomp{#1}{>}{0}{\wire[d][#1][]{q}}{}%
  \ifnumcomp{#1}{<}{0}{\wire[u][-#1][]{q}}{}%
}

\title{\fontsize{14pt}{20pt}\selectfont\textbf{
	A Demonstration of\\
	Quantum Circuit Implementation for Obstacle Flow\\
	Using Carleman-Linearized Lattice Boltzmann Method
}}

\author{\fontsize{12pt}{0}\selectfont
	Kazumasa Ueno,${}^{1,2}$\footnote{
		\url{kazumasa-e67@eps.s.u-tokyo.ac.jp}
	}
	\ \ 
	Keita Kanno,${}^{1}$ 
	\ and\ \ 
	Yasunori Lee${}^{1}$ 
}

\date{\fontsize{11pt}{12pt}\selectfont\textit{
	\setlength{\tabcolsep}{2pt}
	\begin{tabular}{ll}
		${}^1$ & QunaSys, Bunkyo, Tokyo 113-0001, Japan\\[8pt]
		${}^2$ & Department of Earth and Planetary Science, \\
		& The University of Tokyo, Bunkyo, Tokyo 113-0033, Japan
	\end{tabular}
}}

\begin{document}

\begin{titlepage}

\maketitle

\begin{abstract}
Fluid simulations, especially at high Reynolds numbers, are computationally expensive on classical computers, making them promising application targets for quantum computing. Recent studies have combined the lattice Boltzmann method (LBM)
with Carleman linearization to design quantum algorithms for computational fluid dynamics (CFD). However, practical quantum-circuit implementations of these algorithms that incorporate \emph{non-periodic} boundary conditions have not been fully explored.
In this work, we implement a quantum algorithm for two-dimensional linearized fluid flow around an obstacle, using block-encoding of the linear-system matrix and quantum singular value transformation (QSVT) to solve it.
Inflow, outflow, and no-slip boundary conditions are formulated as sparse matrix operations and efficiently embedded into quantum circuits using index-value encoding. We demonstrate logarithmic scaling of the required numbers of qubits and gates with respect to the number of lattice points, suggesting the potential feasibility of quantum-computational fluid dynamics simulations.
\end{abstract}

\thispagestyle{empty}

\renewcommand*{\thefootnote}{\fnsymbol{footnote}}

\end{titlepage}

\newpage

\renewcommand{\baselinestretch}{0.9}
\tableofcontents
\thispagestyle{empty}

\newpage

\renewcommand{\baselinestretch}{1.3}
\setcounter{footnote}{0}
\setcounter{page}{1}

\section{Introduction}\label{sec:intro} 

Research on the application of quantum computing to classical problems has become increasingly active in recent years.
A prominent target is computational fluid dynamics (CFD), which is indispensable in many fields, ranging from automotive engineering to weather forecasting.
Since high-fidelity numerical simulations of fluid flows on classical computers, especially at high Reynolds numbers, are prohibitively costly,
quantum computing holds promise for breakthroughs in this area.

Fluid flows are typically governed by nonlinear partial differential equations (PDEs), such as the Navier--Stokes equations.
However, directly solving these equations is not the only way to simulate them.
One alternative way is to use the \textit{lattice Boltzmann method} (LBM), 
which is a mesoscopic method that represents fluid states by a distribution function $f_q(\bm{x},t)$ of fictitious particles moving on a discrete grid with discrete velocities,
and models the dynamics by solving the Boltzmann kinetic equation in $f_q(\bm{x},t)$. 
Although the Boltzmann equation is still nonlinear,
the LBM is notable for its structural feature that is advantageous for quantum implementations: it separates nonlinearity and nonlocality.
Specifically, the collision step (relaxation toward equilibrium) is nonlinear but local to each lattice site, whereas the streaming step (propagation) is linear and nonlocal.
This makes the nonlinearity in the LBM formulation more amenable to \textit{Carleman linearization} than that in the original Navier-Stokes formulation.

For quantum algorithms for solving time evolution equations, there are two main approaches.
The first approach is to implement the quantum circuit corresponding to a single-time-step evolution
and apply it repeatedly to the quantum state.
For the LBM case, the collision and streaming operators can be \textit{block-encoded} into quantum circuits
based on their representation as linear combinations of unitaries (LCU) \cite{Mezzacapo2015-hw, Budinski2021-iy, Sanavio2024-je} (while costly)
or their structure \cite{Sanavio2025-gn} (with the help of linearization), for instance.
However, this approach in general suffers from exponential decay in the amplitude of the target quantum state with respect to the number of time steps, which poses a significant challenge.

This partly motivates the second approach, which embeds the full time evolution into a single linear system
and solves it all at once using a quantum linear system algorithm (QLSA).
For the LBM case, the discrete-time lattice Boltzmann equation (LBE) can be embedded into a linear system by Carleman linearization,
and the matrix to be inverted can be block-encoded into quantum circuits based on its structure \cite{Li2025-ox, Penuel2024-cl}.

A notable recent contribution in this line is the work of \cite{Jennings2025-theory,Jennings2025-apps}, which developed an end-to-end quantum algorithm for nonlinear fluid dynamics 
and achieved bounded quantum advantage under specific conditions, validated on nontrivial incompressible flow benchmarks.

Despite this progress, there are two key aspects which remain to be addressed in more depth.
First, the dependence of the condition number of the linear system on the problem parameters has not been fully characterized.
In the analysis of \cite{Jennings2025-theory}, all the simulation parameters (i.e., the number of lattice points $N$, the number of time steps $N_t$, and the relaxation time $\tau$) are simultaneously fixed as functions of Reynolds number 
through a specific parameter selection, 
and the scaling of the condition number is analyzed primarily with respect to Reynolds number.
This somewhat obscures the independent contributions of $N_t$ and $N$ to the condition number, which is important for identifying the regime in which the quantum algorithm may be advantageous. 
Also, while \cite{Jennings2025-theory} constructed explicit quantum circuits for the collision and streaming operators in the periodic setting and 
\cite{Jennings2025-apps} extended the algorithm to incorporate nontrivial boundary conditions at the block-encoding level, 
a complete gate-level implementation of the block-encoding of system matrices that include realistic boundary conditions (e.g., inflow, outflow, and no-slip walls) and obstacle geometries
is still lacking.

In this work, we take a step toward filling these gaps by developing a quantum algorithm for simulating flows around an obstacle using first-order Carleman-linearized LBM and a QLSA based on quantum singular value transformation (QSVT).
Following the discrete-time formulation of the LBE, we express the full time-discretized update (collision and streaming together with boundary conditions) as a sparse matrix acting on the vector of distribution functions.
The resulting linear system is solved via matrix inversion based on QSVT.
We analyze the condition number of the linear system and identify that it is primarily governed by the number of time steps $N_t$, with weak dependence on the spatial resolution.
This finding suggests that the quantum algorithm may be most advantageous for problems requiring high spatial resolution but only moderate temporal extent.
Furthermore, we decompose the block-encoding of this matrix into four oracle quantum circuits ($O_{\mathrm{setBC}}$, $O_{\mathrm{collision}}$, $O_{\mathrm{streaming}}$, $O_{\mathrm{unsetBC}}$) that mirror the collision-streaming factorization of the evolution matrix, and describe each oracle's construction in detail.
We provide explicit gate-level resource estimates (qubit counts and gate counts) for representative obstacle-flow configurations.

\newpage

The rest of the paper is organized as follows:
\begin{itemize}
\item In Section~\ref{sec:methods}, we describe the methods used in this study, including the formulation of boundary conditions and the implementation of the quantum circuits.
\item In Section~\ref{sec:results}, we present the results of our simulations, including the condition number analysis and the required numbers of qubits and gates.
\item In Section~\ref{sec:discussion}, we discuss the implications of our findings and possible future directions.
\item In Section~\ref{sec:conclusion}, we summarize our conclusions.
\end{itemize}

\noindent
The notation used in this paper is summarized in Appendix~\ref{appendix:notations}.

\vfill

\begin{figure}[h]
	\centering
	
	\begin{tikzpicture}
		\node[draw, minimum width=200pt] (a) at (0,0) {
			\begin{tabular}{c}
				lattice Boltzmann eq. (BGK)\\
				\eqref{eq:LBE_BGK}
			\end{tabular}
		};
		\node[draw, minimum width=200pt] (b) at (0,-3.33) {
			\begin{tabular}{c}
				nonlinear vector eq.\\
				\eqref{eq:original_LBE}
			\end{tabular}
		};
		\node[draw, minimum width=200pt, minimum height=45pt] (c) at (0,-6.66) {
			\begin{tabular}{c}
				infinite-dim. linear system\\
        \eqref{eq:carleman_collision}, \eqref{eq:carleman_streaming}
			\end{tabular}
		};
		\node[draw, minimum width=200pt] (d) at (0,-10) {
			\begin{tabular}{c}
				finite-dim. linear system\\
				\eqref{eq:linearized_LBE}
			\end{tabular}
		};
		\node[draw, minimum width=200pt] (e) at (10,-10) {
			\begin{tabular}{c}
				another finite-dim. linear system\\
				\eqref{eq:linear_ode}
			\end{tabular}
		};
		\node[inner sep=5pt] (f) at (10,-12.5) {matrix inversion (QSVT etc.)};
		\draw[-Stealth] (a) -- node[right=5pt]{rewrite} (b);
		\draw[-Stealth] (b) -- node[right=5pt]{linearize} (c);
		\draw[-Stealth] (c) -- node[right=5pt]{truncate} (d);
		\draw[-Stealth] (d) -- node[above=5pt]{
			\begin{tabular}{c}
				for quantum\\[-3pt]
				computation
			\end{tabular}
		} (e);
		\draw[-Stealth] (e) -- node[right=5pt]{} (f);
	\end{tikzpicture}
	
	\caption{
		High-level overview of the quantum lattice Boltzmann method.
	}
\end{figure}
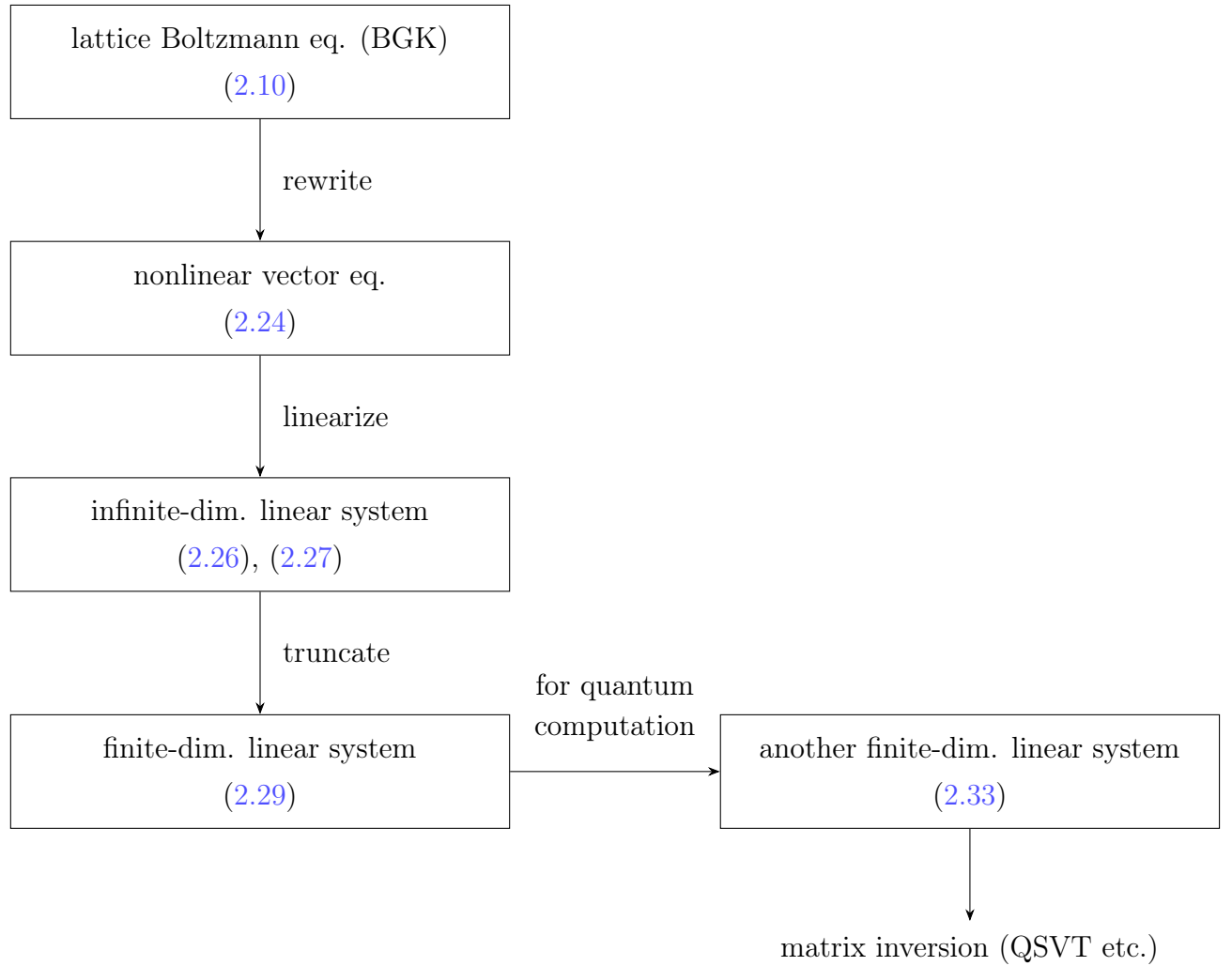

\section{Methods}\label{sec:methods}

\subsection{Lattice Boltzmann method (LBM)}\label{subsec:LBM_review}
Here we briefly summarize the lattice Boltzmann method. For more details, the reader is referred to standard textbooks, e.g., \cite{Kruger2016-lj, Succi2018-yi}. 

\subsubsection{Preliminaries}
The two key dimensionless parameters governing a fluid flow are the Reynolds number
\begin{equation}\label{eq:Re}
	\mathrm{Re} = \frac{U l}{\nu}
\end{equation}
and the Mach number
\begin{equation}\label{eq:Ma}
	\mathrm{Ma} = \frac{U}{c_{\mathrm{s}}}
\end{equation}
where $U$ is the characteristic velocity, $l$ is the characteristic length, $\nu$ is the kinematic viscosity, and $c_{s}$ is the speed of sound.
In the present study, we take $U$ to be the inflow speed and $l$ to be the channel width.

\subsubsection{Description}
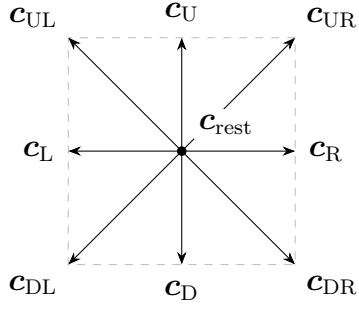
\begin{figure}[h]
	\centering
	\begin{tikzpicture}[scale=1.5]
		\coordinate (0) at (0,0);
		\coordinate (1) at (1,0);
		\coordinate (2) at (1,1);
		\coordinate (3) at (0,1);
		\coordinate (4) at (-1,1);
		\coordinate (5) at (-1,0);
		\coordinate (6) at (-1,-1);
		\coordinate (7) at (0,-1);
		\coordinate (8) at (1,-1);
		\draw[lightgray, dashed] (2) -- (4) -- (6) -- (8) -- cycle;
		\draw[-Stealth] (0) -- (1) node[right=1pt] {$\bm{c}_{\mathrm{R}}$};
		\draw[-Stealth] (0) -- (2) node[above right] {$\bm{c}_{\mathrm{UR}}$};
		\draw[-Stealth] (0) -- (3) node[above=2pt] {$\bm{c}_{\mathrm{U}}$};
		\draw[-Stealth] (0) -- (4) node[above left] {$\bm{c}_{\mathrm{UL}}$};
		\draw[-Stealth] (0) -- (5) node[left=1pt] {$\bm{c}_{\mathrm{L}}$};
		\draw[-Stealth] (0) -- (6) node[below left] {$\bm{c}_{\mathrm{DL}}$};
		\draw[-Stealth] (0) -- (7) node[below=2pt] {$\bm{c}_{\mathrm{D}}$};
		\draw[-Stealth] (0) -- (8) node[below right] {$\bm{c}_{\mathrm{DR}}$};

		\node[above right, fill=white, inner sep=2pt] at (0.1,0.1) {$\bm{c}_{\mathrm{rest}}$};
		\fill (0,0) circle[radius=1.2pt];
	\end{tikzpicture}
	\caption{Unit discrete velocity vectors of the D2Q9 lattice. The vectors $\bm{c}_{\mathrm{R}}$, $\bm{c}_{\mathrm{U}}$, \ldots, $\bm{c}_{\mathrm{DR}}$ correspond to the eight directions connecting a node to its nearest neighbors, and $\bm{c}_{\mathrm{rest}}$ denotes the rest velocity ($|\bm{c}_{\mathrm{rest}}|=0$).}
  \label{fig:D2Q9_lattice}
\end{figure}

In the LBM, fluid states are represented by 
distribution functions $f_q(\bm{x}_n, t)$ at lattice node~$\bm{x}_n$ and time $t$ with discrete velocity $\bm{c}_q$,
from which the macroscopic density $\rho$ and the macroscopic velocity $\bm{u}$ can be obtained as 
\begin{align}
  \rho(\bm{x}_n, t)
  &=
  \sum_q f_q(\bm{x}_n, t),\label{eq:macroscopic_rho}\\
  \rho \bm{u}(\bm{x}_n, t)
  &=
  \sum_q f_q(\bm{x}_n, t) \bm{c}_q.\label{eq:macroscopic_rho_u}
\end{align}
The discrete velocity set $\{\bm{c}_{q}\}$ 
is chosen so that it satisfies isotropy and moment-matching conditions required to recover the original Navier--Stokes equations.
Such velocity sets in $D$ spatial dimensions with $Q$ discrete velocities are called D$D$Q$Q$ models. 
In this study, we consider an $N_x \times N_y$ square lattice and use the D2Q9 model whose discrete velocity vectors and associated weights are shown in Fig.~\ref{fig:D2Q9_lattice} and Table~\ref{tab:D2Q9_weights}, 
and the speed of sound is given by
\begin{equation}
	c_s = \frac{1}{\sqrt{3}} \frac{\Delta x}{\Delta t}
\end{equation}
for the lattice spacing $\Delta x$ and the time step $\Delta t$.

\newpage

\begin{table}[h]
	\centering
	\caption{Weights $w_q$ associated with the discrete velocities in the D2Q9 lattice.}

	\renewcommand*{\arraystretch}{1.3}
	\begin{tabular}{|c|c|c|c|c|c|c|c|c|}
		\hline
		$w_{\mathrm{R}}$ & $w_{\mathrm{U}}$ & $w_{\mathrm{L}}$ & $w_{\mathrm{D}}$ & $w_{\mathrm{UR}}$ & $w_{\mathrm{UL}}$ & $w_{\mathrm{DL}}$ & $w_{\mathrm{DR}}$ & $w_{\mathrm{rest}}$\\
		\hline
		\multicolumn{4}{|c|}{$\frac{1}{9}$} & \multicolumn{4}{c|}{$\frac{1}{36}$} & $\frac{4}{9}$\\
		\hline
	\end{tabular}
  \label{tab:D2Q9_weights}
\end{table}

The evolution of $f_q$ in a time step $\Delta t$ is described by the lattice Boltzmann equation (LBE)
\begin{equation}\label{eq:LBE_general}
  f_q(\bm{x}_n + \bm{c}_q \Delta t, t+\Delta t)
  =
  f_q(\bm{x}_n, t)
  +
  \Omega_q(\bm{x}_n, t),
\end{equation}
where 
$\Omega_q$ is a collision operator.
A widely used choice for $\Omega_q$ is the Bhatnagar--Gross--Krook (BGK) collision operator~\cite{Bhatnagar1954-bgk}
\begin{equation}\label{eq:BGK_operator}
  \Omega_q(\bm{x}_n, t)
  =
  -\frac{\Delta t}{\tau}
  \Big[
    f_q(\bm{x}_n, t) - f_q^{\text{eq.}}(\bm{x}_n, t)
  \Big],
\end{equation}
where 
$f_q^{\text{eq.}}$ is the equilibrium distribution function and $\tau$ is a parameter which determines the rate of relaxation towards it, 
related to $\nu$ as
\begin{equation}
	\nu =	c_{\text{s}}^2 \left( \tau - \frac{\Delta t}{2} \right)
\end{equation}
(as long as $\mathrm{Ma} \ll 1$ and $\mathrm{Ma} \ll \mathrm{Re}$). 
Therefore, given a target value of $\mathrm{Re}$, the relaxation time $\tau$ is fixed as
\begin{equation}\label{eq:tau_from_Re}
  \tau = \frac{1}{c_{\mathrm{s}}^2}\frac{Ul}{\mathrm{Re}} + \frac{\Delta t}{2}.
\end{equation}
In \emph{lattice units} where the lattice spacing $\Delta x = 1$ and the time step $\Delta t = 1$, adopting BGK collision operator reduces the LBE to 
\begin{equation}\label{eq:LBE_BGK}
  f_q(\bm{x}_n + \bm{c}_q, t+1)
  =
  f_q(\bm{x}_n, t)
  -
  \frac{1}{\tau}
  \Big[
    f_q(\bm{x}_n, t) - f_q^{\text{eq.}}(\bm{x}_n, t)
  \Big].
\end{equation}

The evolution in classical LBM is typically implemented in two steps:
the \textit{collision} step first updates the distribution function locally as
\begin{equation}\label{eq:collision_step}
  f_q^*(\bm{x}_n, t)
  =
  f_q(\bm{x}_n, t)
  -
  \frac{1}{\tau}
  \Big[
    f_q(\bm{x}_n, t) - f_q^{\text{eq.}}(\bm{x}_n, t)
  \Big],
\end{equation}
and then the \textit{streaming} step advects $f_q^*$ to neighboring nodes as
\begin{equation}\label{eq:streaming_step}
  f_q(\bm{x}_n + \bm{c}_q, t+1)
  =
  f_q^*(\bm{x}_n, t).
\end{equation}

Hereafter, we work in the lattice units.
Note that the equilibrium distribution function is given by low-order terms in the expansion of the Maxwell--Boltzmann distribution,
which can be written in the lattice units as
\begin{equation}\label{eq:equilibrium_distribution}
  f_q^{\text{eq.}}(\bm{x}_n, t)
  =
  w_q \,\rho(\bm{x}_n, t)
  \left(
    1
    + \frac{\bm{u}(\bm{x}_n, t) \cdot \bm{c}_q}{c_{\text{s}}^2}
    + \frac{(\bm{u}(\bm{x}_n, t) \cdot \bm{c}_q)^2}{2 c_{\text{s}}^4}
    - \frac{\bm{u}(\bm{x}_n, t) \cdot \bm{u}(\bm{x}_n, t)}{2 c_{\text{s}}^2}
  \right).
\end{equation}

\subsubsection{Implementation of boundary conditions}
Periodic boundary conditions are straightforward to implement in LBM: during the streaming step, distribution functions that leave one side of the domain enter from the opposite side.
In contrast, non-periodic boundaries require additional treatment.

\paragraph{No-slip stationary walls.}
We use the bounce-back scheme, in which distributions impinging on the wall are reflected in the opposite directions.
For example, for a wall located halfway between lattice nodes on the bottom boundary, the bounce-back condition is
\begin{equation}\label{eq:bounce_back}
    \begin{aligned}
        f_{\mathrm{U}}(\bm{x}_n, t+1)
        &=
        f_{\mathrm{D}}^*(\bm{x}_n, t), \\
        f_{\mathrm{UR}}(\bm{x}_n, t+1)
        &=
        f_{\mathrm{DL}}^*(\bm{x}_n, t), \text{and}\\
        f_{\mathrm{UL}}(\bm{x}_n, t+1)
        &=
        f_{\mathrm{DR}}^*(\bm{x}_n, t).
    \end{aligned}
\end{equation}

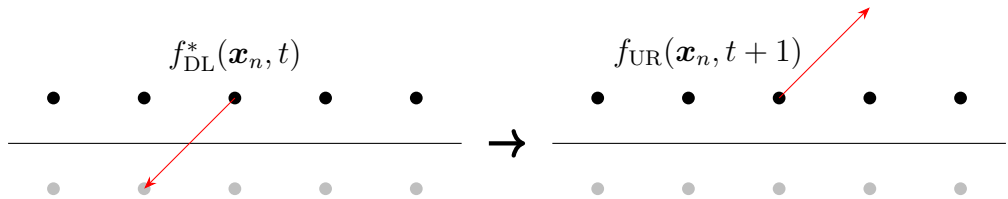
\begin{figure}[h]
	\centering
	\begin{tikzpicture}[scale=1.2]
		\foreach \x in {0,...,4} {
			\fill[lightgray] (\x,0) circle[radius=2pt];
			\fill[]	(\x,1) circle[radius=2pt];
		}
		\foreach \x in {6,...,10} {
			\fill[lightgray] (\x,0) circle[radius=2pt];
			\fill	(\x,1) circle[radius=2pt];
		}
		\draw (-0.5,0.5) -- (4.5,0.5);
		\draw (5.5,0.5) -- (10.5,0.5);
		\draw[->, ultra thick] (4.8,0.5) -- (5.2,0.5);
		\draw[-Stealth, red] (2,1) -- (1,0);
		\node[above=6pt] at (2,1) {$f_{\mathrm{DL}}^*(\bm{x}_n, t)$};
		\draw[-Stealth, red] (8,1) -- node[black, left=4pt] {$f_{\mathrm{UR}}(\bm{x}_n, t+1)$} (9,2);
	\end{tikzpicture}
	\caption{Halfway bounce-back boundary condition at a stationary wall. 
	}
	\label{fig:bounce_back}
\end{figure}

\paragraph{Inflow.}
We impose a prescribed velocity using a moving-wall-type boundary condition.
For example, let us consider an inflow at the left boundary with velocity $\bm{u}_{\text{in}}$.
In this case, for a boundary located halfway between lattice nodes on the left boundary, we set
\begin{equation}\label{eq:moving_wall}
    \begin{aligned}
        f_{\mathrm{R}}(\bm{x}_n, t+1)
        &=
        f_{\mathrm{L}}^*(\bm{x}_n, t)
        +
        \frac{2 w_{\mathrm{L}} \rho_{\text{in}} \bm{u}_{\text{in}} \cdot \bm{c}_{\mathrm{L}}}{c_{\text{s}}^2}, \\
        f_{\mathrm{UR}}(\bm{x}_n, t+1)
        &=
        f_{\mathrm{DL}}^*(\bm{x}_n, t)
        +
        \frac{2 w_{\mathrm{DL}} \rho_{\text{in}} \bm{u}_{\text{in}} \cdot \bm{c}_{\mathrm{DL}}}{c_{\text{s}}^2}, \text{ and}\\
        f_{\mathrm{DR}}(\bm{x}_n, t+1)
        &=
        f_{\mathrm{UL}}^*(\bm{x}_n, t)
        +
        \frac{2 w_{\mathrm{UL}} \rho_{\text{in}} \bm{u}_{\text{in}} \cdot \bm{c}_{\mathrm{UL}}}{c_{\text{s}}^2},
    \end{aligned}
\end{equation}
where $\rho_{\text{in}}$ is the density at the boundary node. 
Note that Eq.~\eqref{eq:moving_wall} specifies, after the streaming step, only the distribution functions whose velocities point into the interior of the fluid domain (right-pointing at the left boundary). These are the directions that would otherwise remain undefined, since they would require streaming from outside the fluid domain. The remaining directions at the same boundary node are obtained by ordinary streaming from interior nodes.

\paragraph{Outflow.}
We use a simple extrapolation from the interior.\footnote{
	Note that this outflow treatment is not commonly used in classical LBM simulations which use more accurate methods.
	However, it is equivalent to the zeroth-order Neumann boundary condition adopted in \cite{Jennings2025-apps} for the outlet, and is sufficient for the purpose of this study. 
}
For example, let us consider an outflow at the right boundary.
For a boundary located halfway between lattice nodes on the right boundary, we set
\begin{equation}\label{eq:outflow}
    \begin{aligned}
        f_{\mathrm{L}}(\bm{x}_n, t+1)
        &=
        f_{\mathrm{L}}(\bm{x}_n-\bm{c}_{\mathrm{R}}, t+1), \\
        f_{\mathrm{UL}}(\bm{x}_n, t+1)
        &=
        f_{\mathrm{UL}}(\bm{x}_n-\bm{c}_{\mathrm{R}}, t+1), \text{ and}\\
        f_{\mathrm{DL}}(\bm{x}_n, t+1)
        &=
        f_{\mathrm{DL}}(\bm{x}_n-\bm{c}_{\mathrm{R}}, t+1).
    \end{aligned}
\end{equation}
As in the inflow case, Eq.~\eqref{eq:outflow} specifies, after the streaming step, only the distribution functions whose velocities point into the interior of the fluid domain (left-pointing at the right boundary), with the remaining directions at the same boundary node obtained by ordinary streaming from interior nodes.

\newpage

\subsection{Quantum lattice Boltzmann method (QLBM)}\label{subsec:QLBM}
In this section, we describe the quantum lattice Boltzmann method using a quantum linear system algorithm, developed in~\cite{Li2025-ox,Penuel2024-cl,Jennings2025-theory,Jennings2025-apps}.
While we largely adopt the formulation developed by the latter two, 
we depart from their choice of the shifted distribution function $g_q \coloneqq f_q - w_q$\footnote{
	Note that we are using the \textit{shifted} variables from the beginning and
	their $\bar{f}_m$ corresponds to our $f_q$.
} and instead work with the standard distribution function $f_q$ throughout this paper.
This choice is sufficient for the first-order Carleman linearization. 

\subsubsection{LBE in vector matrix form}
Hereafter, we adopt the following notation for the velocity and node indices that appear in the collision-streaming pipeline:
\begin{itemize}
  \item $q, n$: \emph{pre-collision} (input) velocity and node indices.
  \item $q^{*}$: \emph{post-collision} velocity index. Nodes are unchanged by collision.
  \item $q_{\mathrm{out}}, n_{\mathrm{out}}$: \emph{post-streaming} (output) velocity and node indices.
\end{itemize}
Primes $'$ are reserved exclusively for additional summation dummy indices (all on the pre-collision side).

Let us  rewrite the equilibrium distribution function \eqref{eq:equilibrium_distribution} in terms of the distribution functions $f$.
By assuming a weakly compressible flow $(1-\rho \ll 1)$, the inverse density is approximated as
\begin{equation*}
	\frac{1}{\rho}
	=
	(1 - (1-\rho))^{-1}
	\approx
	2 - \rho.
\end{equation*}
Substituting Eqs.~\eqref{eq:macroscopic_rho} and \eqref{eq:macroscopic_rho_u} into Eq.~\eqref{eq:equilibrium_distribution}, the equilibrium distribution function can be expressed as 
\begin{equation}\label{eq:f_eq_expanded}
	\begin{array}{ccccl}
		&& \multicolumn{3}{l}{f_{q^*}^{\text{eq.}}(\bm{x}_n, t)}\\[12pt]
		& \approx &
		w_{q^*} \,\bigg[
		&&
		\displaystyle
		\sum_{q} f_{q}(\bm{x}_n, t)\\
		&&& + &
		\displaystyle
		\sum_{q} f_{q}(\bm{x}_n, t) \frac{(\bm{c}_{q} \cdot \bm{c}_{q^*})}{c_{\text{s}}^2}\\
		&&& + &
		\displaystyle
		\sum_{q,q'} f_{q}(\bm{x}_n, t) f_{q'}(\bm{x}_n, t)
		\bigg(2 - \sum_{q''} f_{q''}(\bm{x}_n, t)\bigg)
		\frac{(\bm{c}_{q} \cdot \bm{c}_{q^*})(\bm{c}_{q'} \cdot \bm{c}_{q^*})}{2c_{\text{s}}^4}\\
		&&& - &
		\displaystyle
		\sum_{q,q'} f_{q}(\bm{x}_n, t) f_{q'}(\bm{x}_n, t)
		\bigg(2 - \sum_{q''} f_{q''}(\bm{x}_n, t)\bigg)
		\frac{(\bm{c}_{q} \cdot \bm{c}_{q'})}{2c_{\text{s}}^2} \qquad \bigg]\\[24pt]
		& = &
		w_{q^*} \,\bigg[
		&&
		\displaystyle
		\sum_{q} f_{q}(\bm{x}_n, t)
		\bigg(
			1 + \frac{(\bm{c}_{q} \cdot \bm{c}_{q^*})}{c_{\text{s}}^2}
		\bigg)\\
		&&& + &
		\displaystyle
		\sum_{q,q'} f_{q}(\bm{x}_n, t) f_{q'}(\bm{x}_n, t)
		\bigg(
			\frac{(\bm{c}_{q} \cdot \bm{c}_{q^*})(\bm{c}_{q'} \cdot \bm{c}_{q^*})}{c_{\text{s}}^4}
			-
			\frac{(\bm{c}_{q} \cdot \bm{c}_{q'})}{c_{\text{s}}^2}
		\bigg)\\
		&&& + &
		\hspace{-7pt}
		\displaystyle
		\sum_{q,q',q''}
		\hspace{-7pt} f_{q}(\bm{x}_n, t) f_{q'}(\bm{x}_n, t) f_{q''}(\bm{x}_n, t)
		\bigg(
			- \frac{(\bm{c}_{q} \cdot \bm{c}_{q^*})(\bm{c}_{q'} \cdot \bm{c}_{q^*})}{2c_{\text{s}}^4}
			+ \frac{(\bm{c}_{q} \cdot \bm{c}_{q'})}{2c_{\text{s}}^2}
		\bigg)\ \bigg].
	\end{array}
\end{equation}

We now vectorize the distribution functions to obtain a matrix-vector form of the discrete LBE.
We define a vector $\bm{f}(t) \in \mathbb{R}^{NQ}$ by stacking $f_q(\bm{x}_n, t)$ as
$f(t)[nQ+q] = f_q(\bm{x}_n, t)$, i.e.
\begin{equation}\label{eq:f_vector}
	\bm{f}(t)
	=
	\left(
		\begin{array}{c}
			f_0(\bm{x}_0,t)\\
			\vdots\\
			f_{Q-1}(\bm{x}_0,t)\\
			f_0(\bm{x}_1,t)\\
			\vdots\\
			f_{Q-1}(\bm{x}_1,t)\\
			\vdots\\
			\vdots\\
			f_{Q-1}(\bm{x}_{N-1},t)\\
		\end{array}
	\right).
\end{equation}
Using the expanded equilibrium distribution~\eqref{eq:f_eq_expanded}, the collision step~\eqref{eq:collision_step} can be written in matrix-vector form as
\begin{equation}\label{eq:collision_matrix_form}
	\bm{f}^{*}(t)
	=
	(I + F_1)\bm{f}(t)
	+
	F_2 \bm{f}(t)^{\otimes 2}
	+
	F_3 \bm{f}(t)^{\otimes 3},
\end{equation}
where $\otimes$ denotes the Kronecker product.
Since the collision operator acts locally at each lattice node, the nonzero elements of $F_1 \in \mathbb{R}^{NQ \times NQ}$, $F_2 \in \mathbb{R}^{NQ \times (NQ)^2}$, and $F_3 \in \mathbb{R}^{NQ \times (NQ)^3}$ are given as follows:
\begin{equation}\label{eq:A_elements}
	\begin{array}{lcl}
		(F_1)_{(nQ+q^*)(nQ+q)}
		& = &
		\displaystyle
		\frac{1}{\tau}\bigg(-\delta_{q^* q}
		+
		w_{q^*} \bigg(1 + \frac{\bm{c}_{q} \cdot \bm{c}_{q^*}}{c_{\text{s}}^2}\bigg)\bigg)\\[16pt]
		(F_2)_{(nQ+q^*)[(nQ+q)NQ + (nQ+q')]}
		& = &
		\displaystyle
		\frac{1}{\tau} w_{q^*}
		\bigg(
			\frac{(\bm{c}_{q} \cdot \bm{c}_{q^*})(\bm{c}_{q'} \cdot \bm{c}_{q^*})}{c_{\text{s}}^4}
			-
			\frac{(\bm{c}_{q} \cdot \bm{c}_{q'})}{c_{\text{s}}^2}
		\bigg)\\[16pt]
		(F_3)_{(nQ+q^*)[(nQ+q)(NQ)^2 + (nQ+q')NQ+ (nQ+q'')]}
		& = &
		\displaystyle
		\frac{1}{\tau} w_{q^*}
		\bigg(
			- \frac{(\bm{c}_{q} \cdot \bm{c}_{q^*})(\bm{c}_{q'} \cdot \bm{c}_{q^*})}{2c_{\text{s}}^4}
			+ \frac{(\bm{c}_{q} \cdot \bm{c}_{q'})}{2c_{\text{s}}^2}
		\bigg).\\[16pt]
	\end{array}
\end{equation}

The streaming step~\eqref{eq:streaming_step} advects the post-collision distribution to neighboring nodes. In matrix-vector form, this is written as
\begin{equation}\label{eq:streaming_matrix_form}
	\bm{f}(t + 1)
	=
	S \bm{f}^{*}(t),
\end{equation}
where $S$ is a permutation matrix.
The matrix $S$ is defined so that $S_{(n_{\mathrm{out}}Q+q_{\mathrm{out}})(nQ+q^*)} = 1$ (and all other entries zero), where the output indices $(n_{\mathrm{out}}, q_{\mathrm{out}})$ depend on the boundary condition at node $\bm{x}_n$ for post-collision direction $q^*$.
Corresponding to the boundary conditions in Sec.~\ref{subsec:LBM_review}, the cases are
\begin{equation}\label{eq:streaming_matrix}
	(\bm{x}_{n_{\mathrm{out}}}, q_{\mathrm{out}}) =
	\begin{cases}
		(\bm{x}_{n} + \bm{c}_{q^*},\, q^*) & \text{(interior/periodic)}\\[4pt]
		(\bm{x}_{n},\, \bar{q^*}) & \text{(bounce-back~\eqref{eq:bounce_back})}\\[4pt]
		(\bm{x}_{n},\, \bar{q^*}) & \text{(inflow~\eqref{eq:moving_wall})}\\[4pt]
		(\bm{x}_{n} + \bm{c}_{q^*} + \bm{c}_{\mathrm{R}},\, q^*) & \text{(outflow~\eqref{eq:outflow})}
	\end{cases}
\end{equation}
where $\bar{q}$ denotes the index of the velocity opposite to $\bm{c}_{q}$. 

In addition to the streaming, the inflow boundary condition~\eqref{eq:moving_wall} introduces a velocity-dependent correction proportional to the local density $\rho_{\text{in}} = \sum_{q} f_{q}(\bm{x}_n, t)$.
In general, this correction is linear in $\bm{f}$ and can be absorbed into the streaming-collision matrix.
However, for the quantum circuit implementation, it is more convenient to approximate $\rho_{\text{in}} \approx 1$ and treat the inflow correction as a constant forcing.
Under this approximation, the inflow boundary is handled by the same bounce-back mechanism as the no-slip walls, and the constant velocity corrections
\begin{equation}\label{eq:inflow_forcing}
	b_{n_{\mathrm{out}}Q+q_{\mathrm{out}}}
	=
	\frac{2 w_{q_{\mathrm{out}}} \bm{u}_{\text{in}} \cdot \bm{c}_{q_{\mathrm{out}}}}{c_{\text{s}}^2}
\end{equation}
are grouped into a forcing vector $\bm{b} \in \mathbb{R}^{NQ}$, with $b_{n_{\mathrm{out}}Q+q_{\mathrm{out}}} = 0$ for nodes not on the inflow boundary or for non-outgoing directions.
We adopt this approximation hereafter.

Combining the collision, boundary correction, and streaming steps, the discrete LBE in matrix-vector form reads
\begin{equation}\label{eq:original_LBE}
	\bm{f}(t+1)
	=
	S\Big[(I + F_1)\bm{f}(t)
	+
	F_2 \bm{f}(t)^{\otimes 2}
	+
	F_3 \bm{f}(t)^{\otimes 3}\Big]
	+
	\bm{b}.
\end{equation}

\subsubsection{Carleman linearization}\label{subsubsec:linearization}
Implementing Eq.~\eqref{eq:original_LBE} on a quantum computer requires a linear formulation.
The approach we adopt here is the Carleman linearization~\cite{Carleman1932-zr}, which embeds the nonlinear recurrence into a linear recurrence on an extended space.
Let us define the (infinite-dimensional) Carleman vector by 
\begin{equation}
	\bm{y}^{(\infty)}(t)
	\coloneqq
	\left(
		\begin{array}{c}
			\bm{f}(t)^{\phantom{\otimes 1}}\\
			\bm{f}(t)^{\otimes 2}\\
			\bm{f}(t)^{\otimes 3}\\
			\vdots\\
		\end{array}
	\right)
\end{equation}
and similarly $\bm{y}^{\ast (\infty)}(t)$ by stacking $\bm{f}^*(t)^{\otimes k}$'s.
Then, the collision step~\eqref{eq:collision_matrix_form} can be represented as a linear map 
\begin{equation}\label{eq:carleman_collision}
	\bm{y}^{\ast (\infty)}(t) = \mathcal{C}^{(\infty)}\, \bm{y}^{(\infty)}(t)
\end{equation}
where the concrete form of the Carleman collision matrix $\mathcal{C}^{(\infty)}$ is derived by expanding each $\bm{f}^*(t)$ on the left-hand side.
Similarly, the streaming step~\eqref{eq:streaming_matrix_form} and the constant forcing extend to the Carleman vector as
\begin{equation}\label{eq:carleman_streaming}
	\bm{y}^{(\infty)}(t+1) = \mathcal{S}^{(\infty)}\, \bm{y}^{\ast (\infty)}(t) + \bm{b}_C^{(\infty)},
\end{equation}
where $\mathcal{S}^{(\infty)}$ is the block-diagonal streaming matrix
\begin{equation}
	\mathcal{S}^{(\infty)}
	=
	\begin{pmatrix}
		S & & & \\
		& S^{\otimes 2} & & \\
		&& S^{\otimes 3} & \\
		&&& \ddots
	\end{pmatrix},
\end{equation}
and $\bm{b}_C^{(\infty)}$ is the Carleman-extended forcing vector whose $k$-th block collects the terms involving $\bm{b}$ that arise when taking the $k$-fold Kronecker product of $\bm{f}(t+1)$.

In practice, we truncate the Carleman vector at a finite order $N_C$ and reduce it to a $d_C$-dimensional vector $(d_C \coloneqq \sum_{k=1}^{N_C} (NQ)^k)$
\begin{equation*}
	\bm{y}^{(N_C)}(t)
	\coloneqq
	\left(
		\begin{array}{l}
			\bm{f}(t)\\
			\bm{f}(t)^{\otimes 2}\\
			\ \ \vdots\\
			\bm{f}(t)^{\otimes N_C}
		\end{array}
	\right).
\end{equation*}
Correspondingly, the linear recurrence is also truncated as
\begin{equation}\label{eq:linearized_LBE}
	\bm{y}^{(N_C)}(t+1) = \underbrace{\mathcal{S}^{(N_C)}\mathcal{C}^{(N_C)}}_{\eqqcolon A^{(N_C)}}\, \bm{y}^{(N_C)}(t) + \bm{b}_C^{(N_C)}, 
\end{equation}
where the collision-streaming matrix $A^{(N_C)} \in \mathbb{R}^{d_C \times d_C}$. 
Furthermore, to improve the numerical stability of the time evolution, we introduce a step-size parameter $h \in [0, 1]$ and replace the recurrence~\eqref{eq:linearized_LBE} with the interpolated update
\begin{align}\label{eq:h_interpolation}
	\bm{y}^{(N_C)}(t+h)
	& = (1-h)\bm{y}^{(N_C)}(t) + h\big(A^{(N_C)}\bm{y}^{(N_C)}(t) + \bm{b}_C^{(N_C)}\big)\notag\\
	& = \widetilde{A}^{(N_C)}\,\bm{y}^{(N_C)}(t) + h\bm{b}_C^{(N_C)}
\end{align}
where
\begin{equation}
	\widetilde{A}^{(N_C)} \coloneqq (1-h)I + hA^{(N_C)}.
\end{equation}
This interpolation suppresses patchy inflow transients arising in the linearized dynamics.\footnote{The update~\eqref{eq:h_interpolation} advances the state by the fractional step $h$ rather than a full lattice time unit. After $N_t$ updates the physical time elapsed is $T = N_t h$. The choice $h = 1$ recovers the original recurrence~\eqref{eq:linearized_LBE}.}

For the first-order truncation ($N_C = 1$), the Carleman embedding is equivalent to keeping only the linear term in Eq.~\eqref{eq:original_LBE}, and the collision-streaming matrix reduces to
\begin{equation}\label{eq:matrix_A}
	A^{(1)} = S(I + F_1).
\end{equation}
For each node $n$, input velocity index $q$, and post-collision velocity index $q^{*}$, the matrix element
${A^{(1)}}_{(n_{\mathrm{out}}Q+q_{\mathrm{out}})(nQ+q)}$
is determined by the boundary condition type at node $n$, where $(n_{\mathrm{out}}, q_{\mathrm{out}})$ denotes the streaming destination of $(n, q^{*})$ as defined in Eq.~\eqref{eq:streaming_matrix}.

\newpage

\subsubsection{Time evolution and final state idling}
To compute $\bm{y}(N_t h)$ on a quantum computer, we unroll the recurrence \eqref{eq:h_interpolation} into a single linear system following the approach of
\cite{Berry2014-zu, Jennings2025-theory} 
which at the same time amplifies the probability of extracting the final state upon measurement.

The recurrence is first unrolled into $N_t+1$ block rows (\textit{evolution} steps).
Then, to boost the probability of extracting the final state, $(2^W - 1)N_t-1$ additional \emph{idling} steps are appended. 
During the idling phase, the state is simply copied. 
The corresponding linear system is
\begin{equation}\label{eq:linear_ode}
	\underbrace{
		\begin{pmatrix}
			I &  & & & & & \\
			-\widetilde{A}^{(N_C)} & I & & & & & \\
		 & \ddots & \ddots & & & & \\
		 &  & -\widetilde{A}^{(N_C)} & I & & & \\
		 \hline
		 &  &  & -I & I & & \\
		 &  &  &  & \ddots & \ddots & \\
		 &  &  &  &  & -I & I
		\end{pmatrix}
	}_{\eqqcolon L}
	\begin{pmatrix}
		\bm{y}^{(N_C)}(0) \\
		\bm{y}^{(N_C)}(h) \\
		\vdots \\
		\bm{y}^{(N_C)}(N_t h) \\
		\hline
		\bm{y}^{(N_C)}(N_t h) \\
		\vdots \\
		\bm{y}^{(N_C)}(N_t h)
	\end{pmatrix}
	=
	\underbrace{
		\begin{pmatrix}
			\bm{y}^{(N_C)}(0) \\
			h\bm{b}_C^{(N_C)} \\
			\vdots \\
			h\bm{b}_C^{(N_C)} \\
			\hline
			\bm{0} \\
			\vdots \\
			\bm{0}
		\end{pmatrix}
	}_{\eqqcolon \bm{b}_L}.
\end{equation}
Here, $L$ is a $2^WN_t d_C \times 2^WN_t d_C$ block bidiagonal matrix
whose block rows $2$ to $N_t+1$ have $-\widetilde{A}^{(N_C)}$ (evolution) on the subdiagonal, while block rows $N_t+2$ to $2^WN_t$ have $-I$ (idling) on the subdiagonal.
The right-hand side contains the initial condition $\bm{y}^{(N_C)}(0)$ in the first block, the scaled forcing $h\bm{b}_C^{(N_C)}$ in blocks $2$ to $N_t+1$ (evolution), and zeros in the remaining blocks (idling).

Solving this linear system, one obtains a vector $\bm{y} = L^{-1}\bm{b}_L$
whose first $N_t+1$ blocks encode the full time history $\bm{y}^{(N_C)}(0), \bm{y}^{(N_C)}(h), \ldots, \bm{y}^{(N_C)}(N_t h)$, and the remaining $(2^W-1)N_t-1$ blocks all contain the final state $\bm{y}^{(N_C)}(N_t h)$.
As a result, the final state $\bm{y}^{(N_C)}(N_t h)$ occupies a fraction $(2^W-1)/2^W$ of the solution vector $\bm{y}$, which approaches $1$ exponentially with increasing $W$.
When carried out on a quantum computer with sufficiently large $W$, this ensures that $\bm{y}^{(N_C)}(N_t h)$ can be extracted with high probability upon measurement.

\newpage

\subsection{Quantum Circuit}\label{subsec:quantum_circuit}

The linear system~\eqref{eq:linear_ode} can be solved on a quantum computer using a quantum linear system algorithm (QLSA); 
see, e.g., \cite{morales2025quantumlinearsolverssurvey} for a comprehensive survey of QLSAs.
In this section, we describe the quantum-circuit implementation of QLBM using a QLSA based on the quantum singular value transformation (QSVT) \cite{GilyenSuLowWiebe:2018}.

\subsubsection{State vector and velocity encoding}
The vector $\bm{f}$~\eqref{eq:f_vector} is encoded in a quantum state on $n_{N_x} + n_{N_y} + n_Q$ qubits as
\begin{equation}
	\frac{1}{||\bm{f}||}\sum_{n_x,n_y} \sum_{q=0}^{\widetilde{Q}-1} f_q(\bm{x}_{n_x,n_y}) \ket{n_x} \ket{n_y} \ket{q},
\end{equation}
where $n_{N_x} = \lceil \log_2 N_x \rceil$, $n_{N_y} = \lceil \log_2 N_y \rceil$, $n_Q = \lceil \log_2 Q \rceil$, and $\widetilde{Q} = 2^{n_Q}$. 
The coefficients corresponding to out-of-range indices, namely those not included in the velocity set, are set to zero.
For simplicity, we only consider the case where $N_x$ and $N_y$ are powers of two.

To encode the D2Q9 velocities on $n_Q = 4$ qubits, we use a binary encoding in which the lower two bits $q[1\!:\!0]$ represent the $x$-component and the upper two bits $q[3\!:\!2]$ represent the $y$-component: 
\begin{equation}\label{eq:velocity_encoding}
	q[1\!:\!0] =
	\begin{cases}
		00_{2} & c_x = 0,\\
		01_{2} & c_x = -1,\\
		10_{2} & c_x = +1,
	\end{cases}
	\qquad
	q[3\!:\!2] =
	\begin{cases}
		00_{2} & c_y = 0,\\
		01_{2} & c_y = -1,\\
		10_{2} & c_y = +1.
	\end{cases}
\end{equation}
For example, the velocity index corresponding to $\bm{c}_{\mathrm{UL}}$ in Fig.~\ref{fig:D2Q9_lattice} is encoded as $q[1\!:\!0] = 01_{2}$ and $q[3\!:\!2] = 10_{2}$, giving $q = 1001_2 = 9$.
The nine active velocity indices under this encoding are $q \in \{0, 1, 2, 4, 5, 6, 8, 9, 10\}$, while $q \in \{3, 7, 11, 12, 13, 14, 15\}$ are non-physical padding indices.
The opposite-direction mapping $q \to \bar{q}$ used in the bounce-back boundary reduces to swapping the two bits in each pair: $q[0] \leftrightarrow q[1]$ for the $x$-component and $q[2] \leftrightarrow q[3]$ for the $y$-component.

\subsubsection{Block-encoding of matrix $A^{(1)}$}
The matrix $A^{(1)}$ \eqref{eq:matrix_A} is block-encoded in a unitary $U_{A^{(1)}}$ as 
\begin{equation}\label{eq:block_encoding_A}
  \big(\!\bra{0^{a}}\otimes I\big)\, U_{A^{(1)}} \big(\ket{0^{a}}\otimes I\big) = A^{(1)}/\alpha_A,
\end{equation}
where $a$ is the number of ancilla qubits and $\alpha_A$ is the subnormalization factor.
Equivalently, for each input basis state $\ket{n_x}\ket{n_y}\ket{q}$ on the system register,
\begin{equation}
  U_{A^{(1)}}\big(\ket{n_x}\ket{n_y}\ket{q}\ket{0^{a}}\big)
  =
  \frac{1}{\alpha_A}
  \sum_{n_{\mathrm{out}},\,q_{\mathrm{out}}}
  {A^{(1)}}_{(n_{\mathrm{out}}Q+q_{\mathrm{out}})(nQ+q)}
  \ket{n_{\mathrm{out},x}}\ket{n_{\mathrm{out},y}}\ket{q_{\mathrm{out}}}\ket{0^{a}}
  +
  \ket{\perp},
\end{equation}
where $\ket{\perp}$ is a state orthogonal to the subspace where ancilla qubits are $\ket{0^a}$.

From the Eqs.~\eqref{eq:A_elements} and \eqref{eq:streaming_matrix}, one can see that
the structure of $A^{(1)}$ has two key properties:
\begin{itemize}
  \item The element value depends only on the velocity pair $(q^{*}, q)$ and the boundary condition type, and does not depend on the node index $n$.
  \item The output node $n_{\mathrm{out}}$ depends only on $(n, q^{*})$ and the boundary condition type, and does not depend on the input velocity $q$.
\end{itemize}
Based on these properties, the block-encoding uses the following ancilla registers encoding
\begin{itemize}
	\item post-collision velocity index $q^{*}$ ($n_Q$ qubits),
	\item boundary condition type $\mathrm{BC}$ ($2$ qubits),
	\item amplitude ($1$ qubit).
\end{itemize}
The block-encoding unitary is decomposed as follows using four oracle quantum circuits: 
\begin{enumerate}
  \item 
  Apply Hadamard gates 
  to prepare a uniform superposition over the $q^{\ast}$ register.
  \item $\bm{O_{\mathrm{setBC}}}$\textbf{:} Compute the boundary-condition type at node $\bm{x}_n$ for the post-collision velocity direction $q^{*}$. 
  \item $\bm{O_{\mathrm{collision}}}$\textbf{:} Encode the information of the collision matrix on the amplitude register. 
  \item $\bm{O_{\mathrm{streaming}}}$\textbf{:} 
  Update the node register from $n$ to $n_{\mathrm{out}}$ and, when applicable, the velocity register from $q^{*}$ to $q_{\mathrm{out}}$. 
  \item $\bm{O_{\mathrm{unsetBC}}}$\textbf{:} Uncompute the $\mathrm{BC}$ register based on the updated node and velocity registers. 
  \item 
  Apply Hadamard gates to the $q^{\ast}$ register.
  \item 
  Apply an $X$ gate to the amplitude register so that the block-encoding condition projects onto $\ket{0}$.
\end{enumerate}

\begin{figure}[htbp]
    \centering
    \begin{tikzpicture}
    	\node[scale=1.0] at (0,0) {
    \begin{quantikz}[row sep={20pt,between origins}, column sep=20pt, transparent]
      \lstick{$\ket{n_x}\ket{n_y}$} & \qwbundle{n_{N_x}\!+n_{N_y}} & \qw & \gate[4]{O_{\mathrm{setBC}}} & \qw & \gate[4]{O_{\mathrm{streaming}}} & \gate[4]{O_{\mathrm{unsetBC}}} & \qw & \qw \\
      \lstick{$\ket{q}$} & \qwbundle{n_Q} & \qw & \linethrough & \gate[4]{O_{\mathrm{collision}}} &  & \qw & \qw & \qw \\
      \lstick{$\ket{0}_{q^{\ast}}$} & \qwbundle{n_Q} & \gate{H^{\otimes n_Q}} & \qw & \qw & \linethrough & \linethrough & \gate{H^{\otimes n_Q}} & \qw \\
      \lstick{$\ket{0}_{\mathrm{BC}}$} & \qwbundle{2} & \qw & \qw & \qw & \qw & \qw & \qw & \qw \\
      \lstick{$\ket{0}_{\mathrm{amp.}}$} & \qw & \qw & \qw & \qw & \qw & \qw & \gate{X} & \qw
    \end{quantikz}
    };
    \end{tikzpicture}
\caption{Quantum circuit for $U_{A^{(1)}}$. 
}
\label{fig:block_encoding_A}
\end{figure}
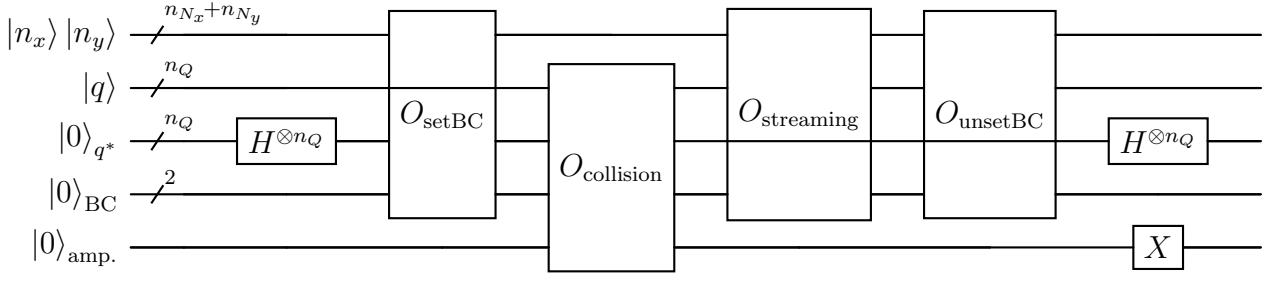

\paragraph{$\bm{O_{\mathrm{setBC}}}$.}
This oracle computes one of the following three boundary-condition types: 
\begin{itemize}
  \item \textbf{Bounce-back} ($\ket{\mathrm{BC}} = \ket{11}$) for
  \begin{itemize}
  \item[$\circ$] nodes at the left boundary ($n_x = 0$) with a left-pointing post-collision velocity.
  \item[$\circ$] nodes at the top ($n_y = N_y - 1$) or bottom ($n_y = 0$) boundaries with velocities pointing into the wall.
  \item[$\circ$] nodes adjacent to the obstacle with velocities pointing into it.
  \end{itemize}
  Since inflow uses the same bounce-back mechanism as no-slip walls, no separate inflow flag is needed.
  \item \textbf{Outflow} ($\ket{\mathrm{BC}} = \ket{10}$) for
  \begin{itemize}
  \item[$\circ$] nodes at the right boundary ($n_x = N_x - 1$) with post-collision velocity being either right-pointing or non-physical ($q^* = 0011_2, 0111_2, 1011_2$).
  \end{itemize}
  \item \textbf{Interior} ($\ket{\mathrm{BC}} = \ket{00}$) otherwise.
\end{itemize}
These conditions are implemented using multi-controlled NOT gates on the $\mathrm{BC}$ register, conditioned on specific bit patterns of $(n_x, n_y, q^{*})$.

\begin{figure}[htbp]
\centering
\begin{quantikz}[row sep={20pt,between origins}, column sep=20pt]
	\lstick{$\ket{n_x}$} & \qwbundle{n_{N_x}}
	& \gate[3, disable auto height, style={rounded corners=20pt}][80pt]{
		\text{bounce-back}
	} & \gate[3, disable auto height, style={rounded corners=20pt}][80pt]{
			\text{outflow}
	} & \qw \\
	\lstick{$\ket{n_y}$} & \qwbundle{n_{N_y}} & \qw & \qw & \qw \\
	\lstick{$\ket{q^{*}}$} & \qwbundle{n_Q} & \ctrl{2} & \ctrl{2} & \qw \\
	\lstick{$\ket{0}_{\mathrm{BC}[0]}$} & \qw & \targ{} & \qw & \qw \\
	\lstick{$\ket{0}_{\mathrm{BC}[1]}$} & \qw & \targ{} & \targ{} & \qw
\end{quantikz}
\caption{Quantum circuit for $O_{\mathrm{setBC}}$.
Each depicted gate actually consists of multiple multi-controlled NOT gates (controlled on each boundary condition).
}
\label{fig:circuit_Oset}
\end{figure}
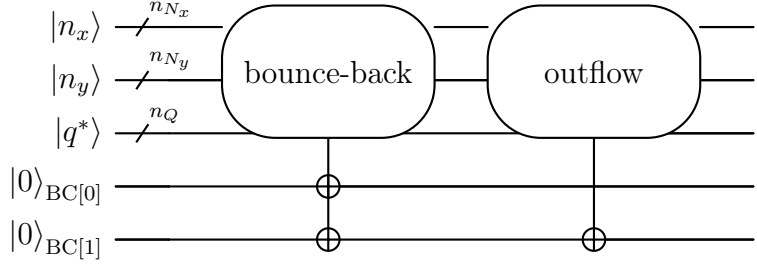

\paragraph{$\bm{O_{\mathrm{collision}}}$.}
This oracle encodes the matrix elements $C_{q^{*} q}$ of the collision matrix $I+F_1$ into the amplitude register via controlled-$R_Y$ gates.
The rotation angle for each pair $(q^{*}, q)$ of active velocity indices (i.e. both in $\{0,1,2,4,5,6,8,9,10\}$) is
\begin{equation}
  \theta_{q^{*} q} = 2\arcsin\!\bigg(\frac{C_{q^{*} q}}{\max_{q^{*}, q}|C_{q^{*} q}|}\bigg),
\end{equation}
where, recalling Eq.~\eqref{eq:A_elements},
\begin{equation}
  C_{q^{*} q} = \Big(1 - \frac{1}{\tau}\Big)\delta_{q^{*} q} + \frac{w_{q^{*}}}{\tau}\Big(1 + \frac{\bm{c}_{q}\cdot\bm{c}_{q^{*}}}{c_{\mathrm{s}}^2}\Big).
\end{equation}
When $C_{q^{*} q} < 0$, a controlled-$Z$ gate needs to be additionally applied to correct the sign.

For the outflow boundary ($\ket{\mathrm{BC}} = \ket{10}$, taken here as the right boundary), the Neumann condition requires that the outward ($c_x = +1$) populations at the boundary be discarded (they have no upstream source) and that the inward ($c_x = -1$) populations be supplied by extrapolation. To this end, additional $R_Y$ gates load the extrapolation values at the non-physical post-collision indices $q^{*} \in \{3, 7, 11\}$ (i.e., $q^{*}[1\!:\!0] = 11_2$), with coefficients corresponding to the left-pointing velocities $q^{*} - 2 \in \{1, 5, 9\}$. The subsequent bit manipulation in $O_{\mathrm{streaming}}$ then routes these values to the inward slot at the right boundary node.

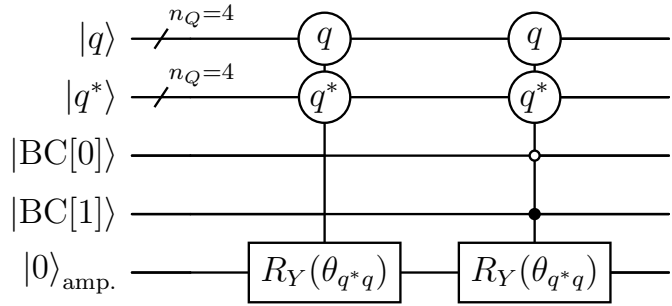
\begin{figure}[htbp]
\centering
	\begin{tikzpicture}
		\node[scale=1.1] at (0,0) {
\begin{quantikz}[row sep={20pt,between origins}, column sep=20pt]
	\lstick{$\ket{q}$} & \qwbundle{n_Q=4}
	& \gate[style={shape=circle, inner sep=-1pt}]{q}\wire[d][4]{q}
	& \gate[style={shape=circle, inner sep=-1pt}]{q}\wire[d][2]{q}
	& \qw \\
	\lstick{$\ket{q^{*}}$} & \qwbundle{n_Q=4}
	& \gate[style={shape=circle, inner sep=-3pt}]{q^{\ast}}
	& \gate[style={shape=circle, inner sep=-3pt}]{q^{\ast}}
	& \qw \\
	\lstick{$\ket{\mathrm{BC}[0]}$} & \qw & \qw & \octrl{1} & \qw \\
	\lstick{$\ket{\mathrm{BC}[1]}$} & \qw & \qw & \ctrl{1} & \qw \\
	\lstick{$\ket{0}_{\mathrm{amp.}}$} & \qw & \gate{R_Y(\theta_{q^{*} q})} & \gate{R_Y(\theta_{q^{*} q})} & \qw
\end{quantikz}
	};
	\end{tikzpicture}
\caption{Quantum circuit for $O_{\mathrm{collision}}$.
Each depicted gate actually consists of multiple multi-controlled rotation gates.} 
\label{fig:circuit_Ocol}
\end{figure}

\paragraph{$\bm{O_{\mathrm{streaming}}}$.}
This oracle first swaps the pre/post-collision velocity registers $\ket{q}$ and $\ket{q^{\ast}}$. 
The streaming operations are then applied to the pre-collision velocity register $\ket{q^{\ast}}_{q}$ (encoding the post-collision velocity $q^{\ast}$ after the swap) and the node register, 
controlled on $\mathrm{BC}$:
\begin{itemize}
  \item \textbf{Interior} ($\ket{\mathrm{BC}} = \ket{00}$): For each velocity component, a controlled adder (incrementer) or subtractor (decrementer) updates the corresponding coordinate register. Specifically, if $q^{*}[0] = 1$ (i.e., $c_x = -1$), subtract 1 from $n_x$; if $q^{*}[1] = 1$ (i.e., $c_x = +1$), add 1 to $n_x$. The $y$-coordinate is updated analogously using $q^{*}[2]$ and $q^{*}[3]$.
  \item \textbf{Bounce-back} ($\ket{\mathrm{BC}} = \ket{11}$): Controlled-SWAP gates exchange $q^{*}[0] \leftrightarrow q^{*}[1]$ (conditioned on $\mathrm{BC}[1] = 1$) and $q^{*}[2] \leftrightarrow q^{*}[3]$ (conditioned on $\mathrm{BC}[0] = 1$), mapping $q^{*}$ to the opposite direction $\bar{q^{*}}$. The node register is unchanged.
  \item \textbf{Outflow} ($\ket{\mathrm{BC}} = \ket{10}$): Building on the extrapolation values pre-loaded by $O_{\mathrm{collision}}$ at $q^{*}[1\!:\!0] = 11_2$, the controlled-SWAP $0\!\leftrightarrow\!1$ (shared with bounce-back, conditioned on $\mathrm{BC}[1]=1$) followed by a controlled-$X$ gate (conditioned on $\mathrm{BC} = 10_2$) exchanging $q^{\ast}[1\!:\!0] = 10_2 \leftrightarrow 11_2$:
  \begin{itemize}
    \item[$\circ$] The extrapolation entries follow $11_2 \xrightarrow{\mathrm{SWAP}} 11_2 \xrightarrow{X_1} 01_2$, becoming the boundary's inward ($c_x = -1$) component for the next time step.
    \item[$\circ$] The $q^{*}[1\!:\!0] = 10_2$ entries follow $10_2 \xrightarrow{\mathrm{SWAP}} 01_2 \xrightarrow{X_1} 11_2$, parked at the non-physical index and thereby decoupled from the LBM dynamics.
  \end{itemize}
  The node register is updated for the $y$-coordinate as in the interior case, while the $x$-coordinate is unchanged. Concretely, the $y$-coordinate adder/subtractor is applied when $\mathrm{BC}[0] = 0$ (i.e., $\mathrm{BC} \in \{00_2, 10_2\}$), while the $x$-coordinate one is applied only when $\mathrm{BC} = 00_2$.
\end{itemize}

\begin{figure}[htbp]
\centering
\begin{quantikz}[row sep={20pt,between origins}, column sep=12pt]
	\lstick{$\ket{n_x}$} & \qwbundle{n_{N_x}} & \qw & \qw & \gate{+ 1} & \gate{- 1} & \qw & \qw & \qw & \qw & \qw & \qw \\
	\lstick{$\ket{n_y}$} & \qwbundle{n_{N_y}} & \qw & \qw & \qw & \qw & \gate{+ 1} & \gate{- 1} & \qw & \qw & \qw & \qw \\
	\lstick{$\ket{q[0]}$} & \qw & \qw & \gate[5][2.1cm]{\mathrm{SWAP}}\gateinput[4]{$q$}\gateoutput[4]{$q^{*}$} & \qw & \ctrl{-2} & \qw & \qw & \swap{1} & \qw & \qw & \qw \\
	\lstick{$\ket{q[1]}$} & \qw & \qw & \qw & \ctrl{-3} & \qw & \qw & \qw & \swap{1} & \qw & \targ{} & \qw \\
	\lstick{$\ket{q[2]}$} & \qw & \qw & \qw & \qw & \qw & \qw & \ctrl{-3} & \qw & \swap{1} & \qw & \qw \\
	\lstick{$\ket{q[3]}$} & \qw & \qw & \qw & \qw & \qw & \ctrl{-4} & \qw & \qw & \swap{1} & \qw & \qw \\
	\lstick{$\ket{q^{*}}$} & \qwbundle{n_Q=4} & \qw & \gateinput{$q^{*}$}\gateoutput{$q$} & \qw & \qw & \qw & \qw & \qw & \qw & \qw & \qw \\
	\lstick{$\ket{\mathrm{BC}[0]}$} & \qw & \qw & \qw & \octrl{-4} & \octrl{-5} & \octrl{-2} & \octrl{-3} & \qw & \ctrl{-2} & \octrl{-4} & \qw \\
	\lstick{$\ket{\mathrm{BC}[1]}$} & \qw & \qw & \qw & \octrl{-1} & \octrl{-1} & \qw & \qw & \ctrl{-5} & \qw & \ctrl{-1} & \qw
\end{quantikz}
\caption{Quantum circuit for $O_{\mathrm{streaming}}$.
}
\label{fig:circuit_Ostr}
\end{figure}

\paragraph{$\bm{O_{\mathrm{unsetBC}}}$.}
This oracle uncomputes the $\mathrm{BC}$ register to $\ket{00}$.
The quantum circuit has the same form as (inverse of) $O_{\mathrm{setBC}}$ except that $O_{\mathrm{unsetBC}}$ needs to control on the \emph{updated} registers.
For the streaming matrix $S$ in Eq.~\eqref{eq:streaming_matrix}, the boundary condition at $(n_{\mathrm{out}}, q_{\mathrm{out}})$ can be determined from $(n_{\mathrm{out}}, q_{\mathrm{out}})$ alone, so the uncomputation is well-defined.
Specifically, the classification in terms of the output registers is:
\begin{itemize}
  \item \textbf{Bounce-back} ($\ket{\mathrm{BC}} = \ket{11}$): Since $O_{\mathrm{streaming}}$ reverses the velocity without shifting the node, the output velocity points away from the boundary. This is for
	\begin{itemize}
	\item[$\circ$] nodes at the left boundary ($n_{\mathrm{out},x} = 0$) with a right-pointing output velocity.
	\item[$\circ$] nodes at the top ($n_{\mathrm{out},y} = N_y - 1$) or bottom ($n_{\mathrm{out},y} = 0$) boundaries with velocities pointing into the interior.
	\item[$\circ$] nodes adjacent to the obstacle with velocities pointing away from it.
	\end{itemize}
  \item \textbf{Outflow} ($\ket{\mathrm{BC}} = \ket{10}$): Since the $n_x$ shift is not applied for outflow, the node remains at the right boundary ($n_{\mathrm{out},x} = N_x - 1$). Nodes at the right boundary with a left-pointing output velocity ($q_{\mathrm{out}}[1\!:\!0] = 01_2$) and non-physical output velocity ($q_{\mathrm{out}} = 0011_2, 0111_2, 1011_2$) are classified as outflow.
  \item \textbf{Interior} ($\ket{\mathrm{BC}} = \ket{00}$): All other output index pairs.
\end{itemize}

\subsubsection{Block-encoding of matrix $L$}
The block-encoding unitary $U_{A^{(1)}}$ is now used to
construct a block-encoding unitary~$U_L$. 
First, let us decompose the block-row index $l \in \{0, 1, \ldots, 2^WN_t - 1\}$ as $l = s \cdot N_t + t$, where $N_t = 2^{n_{N_t}}$ is assumed.
Then, it is encoded on a $(n_t + W)$-qubit register $\ket{t}\ket{s}$ where
\begin{itemize}
  \item $\ket{t}$: time-step register, 
  indexing the time step within each phase.
  \item $\ket{s}$: phase register, 
  distinguishing the evolution $(s=0)$ and idling $(s \geq 1)$ phases.
\end{itemize}

Since $L$ is block bidiagonal, each block row contains at most two nonzero blocks: the diagonal block $I$ and the subdiagonal block $-\widetilde{A}$ or $-I$ depending on the phase. 
Here we block-encode $L$ as a sum of
\begin{itemize}
\item the diagonal $I$'s and the evolution-phase $(s=0)$ subdiagonal $-(1-h)I$'s, and
\item the evolution-phase $(s=0)$ subdiagonal $-hA^{(1)}$'s and the idling-phase $(s \geq 1)$ subdiagonal $-I$'s
\end{itemize}
using an ancilla qubit $\ket{\cdot}_{\mathrm{IN}_L}$ to distinguish between the two.
Since $(1-h) \leq 1$, the maximum absolute value among the individual amplitudes is 
\begin{equation}
	c_L = \max\{1,\, \displaystyle\max_{q^{*},q} h\,|C_{q^{*} q}|\},
\end{equation}
by which each rotation is normalized.
Together with the factors due to the Hadamard gates acting on the $q^{\ast}$ register and 
the summand-indicator $\mathrm{IN}_L$ register
at both ends of the circuit, the overall subnormalization factor of the block-encoding is
\begin{equation}\label{eq:subnorm_L}
  \alpha_L = c_L \cdot (\sqrt{2}^{n_Q+1})^2. 
\end{equation}

The block-encoding of $L$ proceeds as follows:
\begin{enumerate}
  \item 
  Apply Hadamard gates to the $q^{\ast}$ and $\mathrm{IN}_L$ registers.
  \item \textbf{Diagonal term:} Apply a controlled-$R_Y$ gate to the amplitude register encoding the value $+1/c_L$, conditioned on 
  $\ket{0}_{\mathrm{IN}_L}\ket{q^{\ast}_0}_{q^{\ast}}$ for an arbitrarily chosen index $q^{\ast}_0$.
  \item \textbf{Evolution subdiagonal, $-(1-h)I$ term:} Apply a controlled-$R_Y$ gate to the amplitude register encoding the value $-(1-h)/c_L$, conditioned on
	$\ket{0}_{\mathrm{IN}_L}\ket{q^{\ast}_1}_{q^{\ast}}\ket{0}_s$ for an arbitrarily chosen index $q^{\ast}_1$ satisfying $q^{\ast}_1 \neq q^{\ast}_0$.
  \item \textbf{Evolution subdiagonal, $-hA^{(1)}$ term:} Controlled on
	$\ket{1}_{\mathrm{IN}_L}\ket{0}_s$,
  apply the block-encoding $U_{A^{(1)}}$ to the system register, which encodes $hC_{q^{*} q}/c_L$ into the amplitude. 
  \item \textbf{Idling subdiagonal, $-I$ term:} Apply a controlled-$R_Y$ gate to the amplitude register encoding the value $-1/c_L$, conditioned on 
  $\ket{1}_{\mathrm{IN}_L}\ket{q^{\ast}_2}_{q^{\ast}}\ket{s}_s$ for any $s \geq 1$ and for an arbitrarily chosen index $q^{\ast}_2$.
  The last block row ($s = 2^W - 1$, $t = N_t - 1$) is excluded from this term using an auxiliary qubit.
  \item \textbf{Row shift:} Increment the block-row index 
  $l \to l + 1$ to map the subdiagonal column index to the correct row.
  \item 
  Apply Hadamard gates to the $q^{\ast}$ and $\mathrm{IN}_L$ registers.
  \item 
  Apply an $X$ gate to the amplitude register.
\end{enumerate}

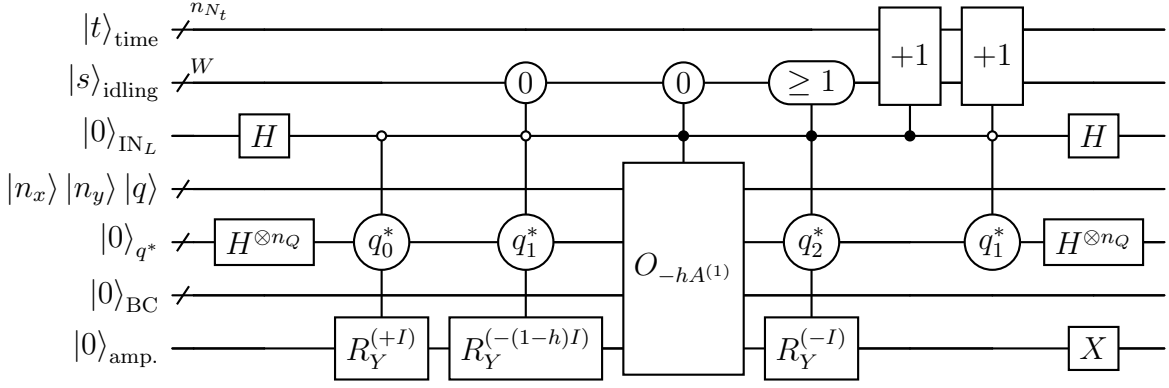
\begin{figure}[h]
    \centering
		\begin{quantikz}[row sep={20pt,between origins}, column sep=8pt]
			\lstick{$\ket{t}_{\text{time}}$} & \qwbundle{n_{N_t}} & \qw & \qw & \qw & \qw & \qw & \gate[2]{+1} & \gate[2]{+1} & \qw & \qw \\
			\lstick{$\ket{s}_{\text{idling}}$} & \qwbundle{W} & \qw & \qw & \hexlctrl{1}{0} & \hexlctrl{1}{0} &
			\gate[disable auto height, style={rounded corners=8pt}][32pt]{\geq 1} \wire[d][1]{q}
			& \qw & \qw & \qw & \qw \\
			\lstick{$\ket{0}_{\mathrm{IN}_L}$} & \qw & \gate{H} & \octrl{2} & \octrl{2} & \ctrl{1} & \ctrl{2} & \ctrl{-1} & \octrl{-1} & \gate{H} & \qw \\
			\lstick{$\ket{n_x}\ket{n_y}\ket{q}$} & \qwbundle{} & \qw & \qw & \qw & \gate[4]{O_{-hA^{(1)}}} & \qw & \qw & \qw & \qw & \qw \\
			\lstick{$\ket{0}_{q^{\ast}}$} & \qwbundle{} & \gate{H^{\otimes n_Q}} & \hexlctrl{2}{q^{\ast}_0} & \hexlctrl{2}{q^{\ast}_1} & \qw & \hexlctrl{2}{q^{\ast}_2} & \qw & \hexlctrl{-2}{q^{\ast}_1} & \gate{H^{\otimes n_Q}} & \qw \\
			\lstick{$\ket{0}_{\mathrm{BC}}$} & \qwbundle{} & \qw & \qw & \qw & \qw & \qw & \qw & \qw & \qw & \qw \\
			\lstick{$\ket{0}_{\mathrm{amp.}}$} & \qw & \qw & \gate{R_Y^{(+I)}} & \gate{R_Y^{(-(1-h)I)}} & \qw & \gate{R_Y^{(-I)}} & \qw & \qw & \gate{X} & \qw
		\end{quantikz}
\caption{Quantum circuit for $U_L$.
Here, $O_{-hA^{(1)}}$ is the inner part of the block-encoding circuit
$U_{A^{(1)}}$ in Fig.~\ref{fig:block_encoding_A}, i.e., the sequence of $O_{\mathrm{setBC}}$, $O_{\mathrm{collision}}$, $O_{\mathrm{streaming}}$, and $O_{\mathrm{unsetBC}}$, 
with $O_{\mathrm{collision}}$ modified to encode $-hC_{q^{\ast} q}/c_L$ in place of $C_{q^{\ast} q}/\max_{q^{\ast}, q}|C_{q^{\ast} q}|$.
The control indices $q^{\ast}_{0}$, $q^{\ast}_{1}$, and $q^{\ast}_{2}$ of the $R_Y$ gates can be chosen arbitrarily as long as $q^{\ast}_0 \neq q^{\ast}_1$.
}
\label{fig:block_encoding_L}
\end{figure}


\subsubsection{QLSA by QSVT}
\begin{figure}[htbp]
  \centering
  \begin{quantikz}[row sep={30pt,between origins}, column sep=10pt]
    \lstick{$\ket{0}_{}$} & \qw & \targ{} & \gate{e^{i\phi_0 Z}} & \targ{} & \qw & \targ{} & \gate{e^{i\phi_1 Z}} & \targ{} & \qw & \ \ldots\ \qw & \targ{} & \gate{e^{i\phi_{\mathrm{deg}\,P} Z}} & \targ{} & \qw \\
    \lstick{$\ket{0^{a}}$} & \qwbundle{} & \hexlctrl{-1}{0} & \qw & \hexlctrl{-1}{0} & \gate[2]{U_L} & \hexlctrl{-1}{0} & \qw & \hexlctrl{-1}{0} & \gate[2]{U_L^{\dagger}} & \ \ldots\ \qw & \hexlctrl{-1}{0} & \qw & \hexlctrl{-1}{0} & \qw \\
    \lstick{$\ket{\bm{b}_{L}}$} & \qwbundle{} & \qw & \qw & \qw & \qw & \qw & \qw & \qw & \qw & \ \ldots\ \qw & \qw & \qw & \qw & \qw
  \end{quantikz}
\caption{A quantum circuit for the QSVT-based QLSA. 
The ancilla register $\ket{0^a}$ includes all ancilla qubits used in the block-encoding of $L$ (i.e., $q^{\ast}$, $\mathrm{IN}_L$, $\mathrm{BC}$, and the amplitude registers in our case).
}
\label{fig:qsvt_circuit}
\end{figure}
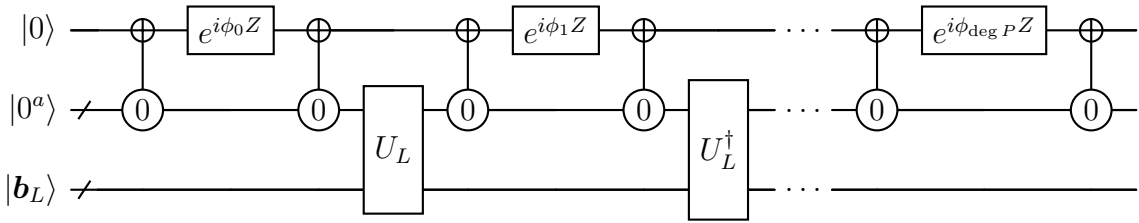

To solve the linear system of equations \eqref{eq:linear_ode}, we employ a QLSA based on QSVT~\cite{GilyenSuLowWiebe:2018}.
For a block-encoded matrix $M$ with the singular value decomposition $V\Sigma W^{\dagger}$,
QSVT by an odd (resp. even) polynomial~$P$ realizes a block-encoding of $P_{\mathrm{SV}}(M) \coloneqq VP(\Sigma)W^{\dagger}$ (resp. $VP(\Sigma) V^{\dagger}$).
A circuit which achieves this consists of alternating applications of the block-encoding unitary and its inverse, interleaved with controlled phase rotations corresponding to the polynomial $P$, 
as shown in Fig.~\ref{fig:qsvt_circuit}.

If one chooses an odd polynomial $P_{\text{inv.}}$ approximating $1/(\kappa_{\mathrm{eff.}}\, x)$ on $[1/\kappa_{\mathrm{eff.}}, 1]$ 
such that the range contains all the singular values of the (subnormalized) block-encoded matrix,
QSVT (approximately) achieves matrix inversion and solves the linear system. 
The degree $d_{\text{QSVT}}$ of the polynomial required to achieve an approximation error of at most $\epsilon_{\text{QSVT}}$ is known to be
\begin{equation}\label{eq:degree_polynomial}
	d_{\text{QSVT}}
  = {\mathcal{O}}(\kappa_{\mathrm{eff.}} \log\bigg(\frac{\kappa_{\mathrm{eff.}}}{\epsilon_{\text{QSVT}}}\bigg)),
\end{equation}
and this determines the number of queries to the block-encoding unitary.\footnote{
	In this work, we used the polynomial introduced in \cite{GriblingKerenidisSzilágyi} and computed the angles based on \cite{BerntsonSünderhauf:2024}.
} 

For the case of QLBM \eqref{eq:linear_ode}, if the approximation parameter $\kappa_{\mathrm{eff.}}$ is chosen so that it satisfies
\begin{equation*}
	\kappa_{\mathrm{eff.}} \geq \frac{1}{\sigma_{\min}(L/\alpha_L)}
\end{equation*}
where the right-hand side accounts for both the condition number of the original linear system and the block-encoding subnormalization factor,
then the range $[1/\kappa_{\mathrm{eff.}},1]$ contains all the singular values of the block-encoded matrix to be inverted by QSVT.
According to \cite[Lem.\,6]{Berry2014-zu}, 
\begin{equation}\label{eq:norm_inverse_L}
  \|L^{-1}\|_2 = {\mathcal{O}}(N_t \kappa_V)
\end{equation}
where $\kappa_V = \|V\|_2\,\|V^{-1}\|_2$ is the condition number of the matrix $V$ that diagonalizes the evolution matrix $I - \mathcal{S}\mathcal{C}$. 
Since the subnormalization factor $\alpha_L$ is independent of $N_t$, it follows that $\kappa_{\mathrm{eff.}} \geq \alpha_L\,\|L^{-1}\|_2 = \mathcal{O}(\alpha_L\, N_t \kappa_V)$.

The output state, upon successful post-selection on the ancilla registers being $\ket{0}$, is proportional to $L^{-1}\ket{\bm{b}_L}$. The success probability of this post-selection is $\Omega(1/\kappa_{\mathrm{eff.}}^2)$. In practice, when one uses quantum amplitude estimation 
to extract observables from the solution state, the estimation procedure internally handles the low success probability at the cost of an additional $\mathcal{O}(\kappa_{\mathrm{eff.}})$ multiplicative overhead in the query complexity.
It is therefore important to estimate $\kappa_{\mathrm{eff.}}$ to evaluate the quantum resources required for the QLBM, which we will investigate in Sec.~\ref{subsec:condition_number}.

\newpage

\section{Results}\label{sec:results}
\begin{figure}[htbp]
  \centering
  \begin{tikzpicture}[scale=0.6]
    \draw[step=1, very thin, gray!50, dashed] (0,0) grid (8,8);

	\draw[very thick] (0,0) -- node[below=3pt] {bounce-back (no-slip)} (8,0);
	\draw[very thick] (0,8) -- (8,8);
	\draw[very thick, dashed] (0,0) -- (0,8);
	\draw[very thick, dashed] (8,0) -- (8,8);

    \fill[black] (2,3) rectangle (3,5);

    \draw[very thick,->] (-1,5.5) -- node[left=18pt]{inflow} (1,5.5);
    \draw[very thick,->] (7,5.5) -- node[right=18pt]{outflow} (9,5.5);
  \end{tikzpicture}
  \caption{Flow configuration for the 2D channel with a rectangular obstacle. 
  	The gray dashed lines are added to indicate the size and position of the obstacle.
  }
  \label{fig:channel_obstacle}
\end{figure}
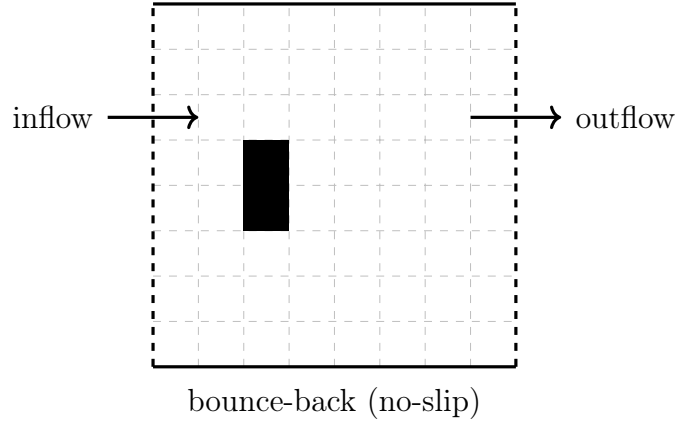

In this study, we consider two-dimensional flow around a rectangular obstacle in a square computational domain as a benchmark problem to evaluate the performance of the proposed QLBM. 
The number of grid points 
are set to powers of two with $N_x = N_y$ for simplicity, 
and an obstacle of size $N_x/8 \times N_y/4$ is placed in the upstream region of the channel, as shown in Fig.~\ref{fig:channel_obstacle}. 

\noindent
The common physical parameters are set as follows:
\begin{center}
	\begin{tabular}{lcl}
		step-size parameter & : & $h = 0.5$\\
		Carleman truncation order & : & $N_C = 1$\\
		idling phase parameter & : & $W = 1$\\
		Reynolds number & : & $\mathrm{Re} = 1$ (based on the channel width)\\
		Mach number of the inflow velocity & : & $\mathrm{Ma} = 0.01$
	\end{tabular}
\end{center}
The initial condition is the fluid at rest (zero velocity everywhere, i.e. $f_q(\bm{x}_n,0) \propto w_q$).


\subsection{Effective condition number of the matrix $L$}\label{subsec:condition_number}

We first compute the condition number and the maximum/minimum singular values of $L$ for various grid
sizes $N_x\,(=N_y)$ and time-step counts $N_t$. 
We take the channel width $N_y$ (in lattice units) as the characteristic length $l$ and $\mathrm{Ma}\cdot c_{\mathrm{s}}$ as the characteristic velocity $U$. The relaxation time $\tau$ is then determined by Eq.~\eqref{eq:tau_from_Re}.

\newpage

\begin{figure}[htbp]
  \centering
  \includegraphics[width=\linewidth]{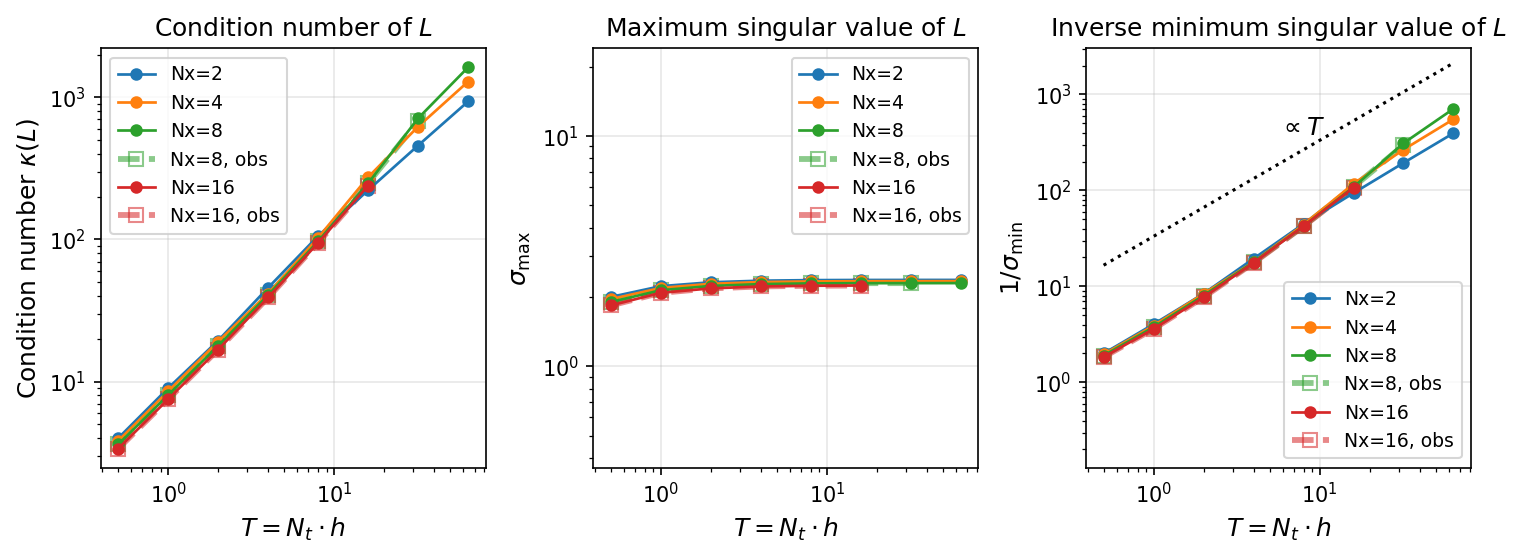}
  \caption{Spectral properties of the matrix $L$ as a function of the total simulation time $T = N_t \cdot h$, for various grid sizes $N_x \,(= N_y)$. 
    }
  \label{fig:condition_number_L}
\end{figure}

The results are shown in Fig.~\ref{fig:condition_number_L}.
As the total simulation time $T = N_t h$ increases, the condition number of $L$ grows approximately linearly with $T$, which is due to the decrease of the minimum singular value. 
The growth of $\|L^{-1}\|$ with $T$ is consistent with the bound in Eq.~\eqref{eq:norm_inverse_L}, $\|L^{-1}\| = O(N_t \kappa_V)$, since $N_t$ is proportional to $T$ and $\kappa_V$ is independent of $T$.
Furthermore, neither the spatial resolution $N_x$ nor the presence of the
obstacle significantly affects these spectral properties, confirming that $T$
is the dominant driver of $\kappa_{\mathrm{eff.}}$.
The effective condition number for a given problem instance is then obtained by
multiplying $1/\sigma_{\min}(L)$ by the subnormalization factor $\alpha_L$.

\subsection{Simulation results of flow around an obstacle}\label{subsec:simulation_results}
We validate the proposed QLBM through two complementary approaches. First, we perform a state-vector simulation of the full quantum circuit on a small problem instance and compare it with a classical time-stepping simulation to confirm that the quantum algorithm correctly reproduces the linearized LBM dynamics. Second, we employ the Clenshaw algorithm to classically emulate the QSVT-based polynomial transformation. The state-vector simulation of the full quantum circuit requires simulating all qubits including the ancilla qubits used for block-encoding, which severely limits the tractable problem size. The Clenshaw algorithm operates directly on the matrix $L$ without the ancilla overhead, enabling the same QSVT polynomial to be evaluated on significantly larger problem instances. This allows us to study larger $N_t$ and the dependence of the solution accuracy on the polynomial degree $d_{\mathrm{QSVT}}$ and the effective condition number $\kappa_{\mathrm{QSVT}}$.

\newpage

\subsubsection{Quantum circuit simulation}\label{subsubsec:quantum_circuit_sim}
\begin{figure}[htbp]
  \centering
  \includegraphics[width=\linewidth]{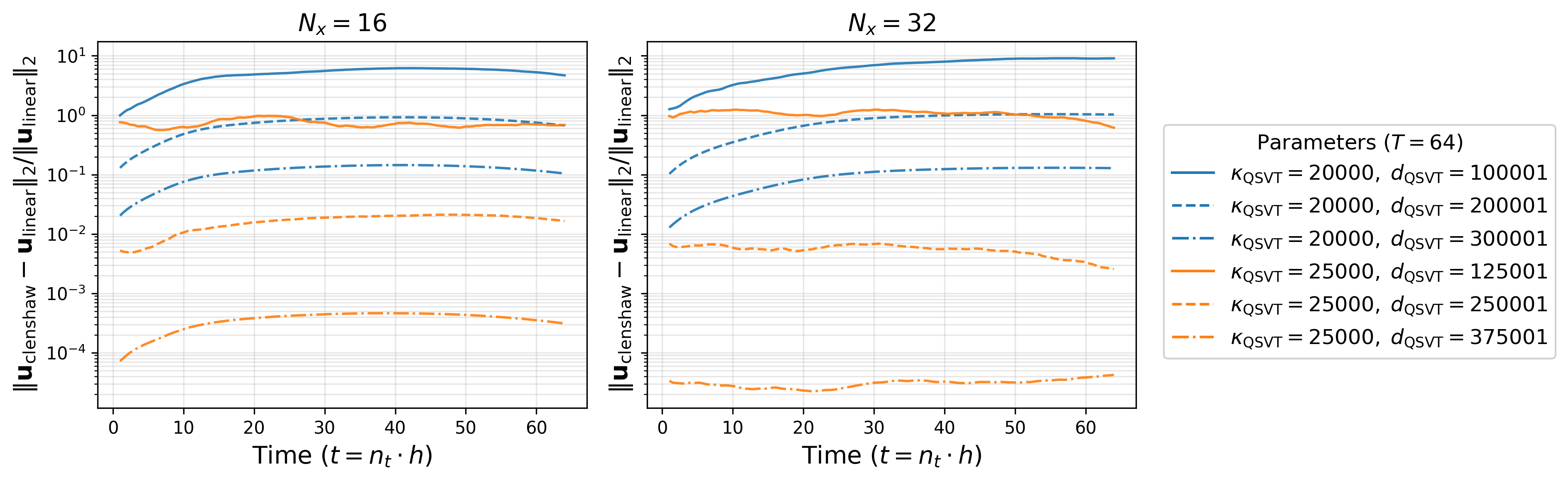}
  \caption{%
    Accuracy of the Clenshaw-based solution with respect to
    the linear reference as a function of intermediate time $t$,
    with total simulation time $T = 64$.
    }
  \label{fig:clenshaw_error_vs_Nt}
\end{figure}

We perform a state-vector simulation of the full quantum circuit to verify that the quantum algorithm correctly reproduces the expected linearized LBM dynamics. Due to the exponential cost of state-vector simulation, we consider a small-scale problem with 
\begin{itemize}
\item $N_x = N_y = 8$
\vspace{-4pt}
\item $N_t = 32$ $(T=16)$
\vspace{-4pt}
\item $\kappa_{\mathrm{QSVT}} = 3500$\,: chosen to exceed $1/\sigma_{\min}(L) \times \alpha_L \approx 3488$, where $1/\sigma_{\min}(L) \approx 109$ is read from Fig.~\ref{fig:condition_number_L} and $\alpha_L = 1 \cdot 2^5 = 32$ is the subnormalization factor of Eq.~\eqref{eq:subnorm_L}.
\vspace{-4pt}
\item $d_{\mathrm{QSVT}} = 35001$\,: chosen based on Fig.~\ref{fig:clenshaw_error_vs_Nt} (presented later) to suppress the polynomial approximation error below $1\%$.
\end{itemize}

\begin{figure}[htbp]
  \centering
  \includegraphics[width=\linewidth]{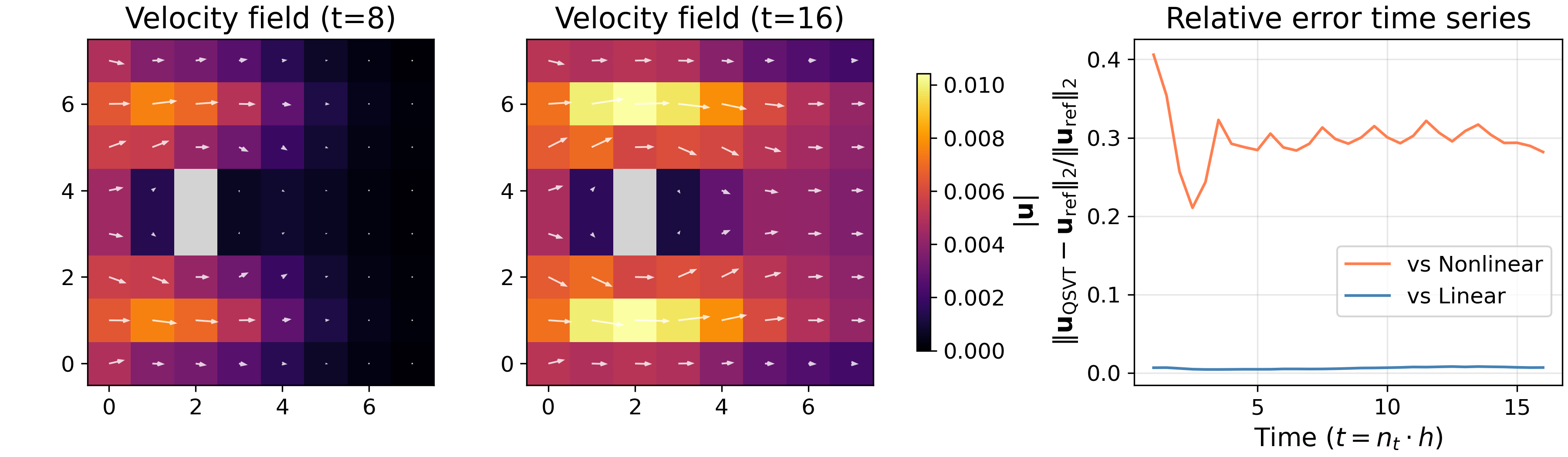}
  \caption{State-vector simulation results of the QLBM\\
  ($N_x = N_y = 8$, $N_t = 32$ ($T = 16$), $\kappa_{\mathrm{QSVT}} = 3500$, $d_{\mathrm{QSVT}} = 35001$).}
  \label{fig:quantum_results}
\end{figure}

We compare the QLBM result with two classical references: (i) a classical time-stepping simulation of the linearized LBM (\textit{linear reference}), and (ii) a classical time-stepping simulation of the original nonlinear LBM (\textit{nonlinear reference}). Figure~\ref{fig:quantum_results} shows the velocity fields at an intermediate time $t = 8$ and at the final time $t = T = 16$, and the 2-norm of the error as a function of time $t$.
The QLBM result is in good agreement with the linear reference at all times, confirming that the quantum circuit correctly implements the intended linear system solver. Notably, the error of QLBM with respect to the linear reference remains well below that with respect to the nonlinear reference, suggesting that, in this particular case, the numerical error introduced by the QSVT-based polynomial approximation is negligible compared to the truncation error in the Carleman linearization. 

\subsubsection{Classical emulation via the Clenshaw algorithm}\label{subsubsec:clenshaw}
While the quantum circuit simulation validates the correctness of the QLBM, state-vector simulation of the full circuit requires exponential memory in the total number of qubits, including the numerous ancilla qubits introduced by the block-encoding construction. 
To circumvent this limitation and extend our analysis to problems with larger scale,
we employ the Clenshaw algorithm~\cite{Clenshaw1955-iy} to directly evaluate the polynomial transformation on the matrix $L$ which is supposed to be realized by the QSVT quantum circuit.\footnote{
Note that an actual quantum circuit implementation additionally involves converting polynomial coefficients to QSVT phase angles and synthesizing the resulting rotation gates into Clifford$+\mathrm{T}$ gate sequences, each of which introduces its own approximation errors. The results presented here should therefore be understood as evaluations of the ideal polynomial transformation, not as predictions of the actual output of a physical quantum device.
}

In the QSVT-based approach, the solution vector of the linear system $\bm{y}=L^{-1
}\bm{b}_L$ is obtained by applying a polynomial transformation to the singular values of $L/\alpha_L$:
\begin{equation}\label{eq:qsvt_solution}
  \bm{y} \approx P_{\mathrm{SV}}[L/\alpha_L]\, \bm{b}_L = \sum_{j=0}^{d_{\mathrm{QSVT}}} a_j\, T_{j\, \mathrm{SV}}\!\left[\frac{L}{\alpha_L}\right] \bm{b}_L,
\end{equation}
where $T_j$ is the Chebyshev polynomial of degree $j$ and $\{a_j\}$ are the expansion coefficients of the matrix-inversion polynomial $P(x) \approx 1/(\kappa_{\mathrm{QSVT}}\, x)$ in the Chebyshev basis. 

The Clenshaw algorithm evaluates the sum in Eq.~\eqref{eq:qsvt_solution} efficiently using the three-term recurrence relation of the Chebyshev polynomials. Since $L$ is non-Hermitian, we work with the extended Hermitian matrix
\begin{equation}\label{eq:extended_H}
  H =
  \begin{pmatrix}
    0 & L/\alpha_L \\
    L^{\dagger}/\alpha_L & 0
  \end{pmatrix}
\end{equation}
and the extended vector $\bm{x} = (\bm{b}_L,\, \bm{0})^{\mathsf{T}}$.
The Clenshaw recurrence defines a sequence of vectors
\begin{equation}\label{eq:clenshaw_recurrence}
	\bm{y}_{j} = 
	\begin{cases}
		\bm{0} & (j = d_{\mathrm{QSVT}}+1)\\
		a_{d_{\mathrm{QSVT}}}\, \bm{x} & (j = d_{\mathrm{QSVT}})\\
		2H\, \bm{y}_{j+1} + a_j\, \bm{x} - \bm{y}_{j+2} & (\text{otherwise})\\
	\end{cases}
\end{equation}
and the result is given by $\sum_{j=0}^{d_{\mathrm{QSVT}}} a_j\, T_j(H)\, \bm{x} = \bm{y}_0 - H\, \bm{y}_1$, whose lower half yields $P_{\mathrm{SV}}[L/\alpha_L]\, \bm{b}_L$. Naively evaluating a degree-$d_{\mathrm{QSVT}}$ matrix polynomial would require $d_{\mathrm{QSVT}}$ matrix-matrix products, but the Clenshaw recurrence avoids this entirely. Each iteration requires only a single sparse matrix-vector product with $H$ and vector additions, keeping the cost linear in $d_{\mathrm{QSVT}}$ and tractable even for large polynomial degrees.

\paragraph{Validation against the quantum circuit simulation.}
We first confirm that the Clenshaw algorithm reproduces the QSVT quantum circuit simulation results.
Using the same problem size and parameters as in Sec.~\ref{subsubsec:quantum_circuit_sim} ($N_x = N_y = 8$, $N_t = 32$), we compared the QSVT and Clenshaw outputs for several combinations of $\kappa_{\mathrm{QSVT}}$ and $d_{\mathrm{QSVT}}$, including deliberately under-resolved settings where $\kappa_{\mathrm{QSVT}}$ lies below the estimated $\alpha_L/\sigma_{\min}(L)$.

In all cases, the Clenshaw-based solution agrees with the QSVT output to within a relative $\ell_2$ error of $2 \times 10^{-6}$, even when the QSVT solution itself exhibits a relative error of several percent with respect to the linear LBM reference (e.g.\ $\kappa_{\mathrm{QSVT}} = 3000$, $d_{\mathrm{QSVT}} = 15001$).
We note that this $\sim 10^{-6}$ residual, while far above machine precision, is consistent with the error introduced when converting polynomial coefficients to QSVT phase angles, which the Clenshaw algorithm bypasses by evaluating the polynomial directly from the coefficients.
Since this level of discrepancy is negligible compared to the polynomial approximation error itself, the Clenshaw algorithm can be used as a proxy for the QSVT quantum algorithm, 
in the following larger-scale studies.

\begin{table}[h]
  \centering
  \caption{Relative $\ell_2$ errors between the outputs ($N_x = N_y = 8$, $N_t = 32$, $h = 0.5$).
	}
  \label{tab:clenshaw_validation}
  \begin{tabular}{cccc}
    \hline
    $\kappa_{\mathrm{QSVT}}$ & $d_{\mathrm{QSVT}}$ & QSVT vs. Clenshaw & QSVT vs. linear LBM \\
    \hline
    3000 & 15001 & $2.1 \times 10^{-7}$ & $4.8 \times 10^{-2}$ \\
    3000 & 30001 & $8.2 \times 10^{-7}$ & $4.0 \times 10^{-3}$ \\
    3500 & 17501 & $3.6 \times 10^{-7}$ & $8.5 \times 10^{-3}$ \\
    3500 & 35001 & $1.7 \times 10^{-6}$ & $4.8 \times 10^{-5}$ \\
    \hline
  \end{tabular}
\end{table}


\paragraph{Larger-scale simulation.}
Having validated the Clenshaw algorithm, we apply it to larger problem sizes
$N_x = N_y = 16, 32$ with total simulation times
$T = 16,\, 32,\, 64$. 
Here $T = N_t h$ is the \emph{total} (maximum) simulation time,
whereas $t = n_t h$ $(n_t = 0,1,\ldots,N_t)$ denotes the
\emph{intermediate} time at step $n_t$. 
For each $T$, we choose two values of the condition number parameter
$\kappa_{\mathrm{QSVT}}$: a \emph{low} value, set below
$\alpha_L/\sigma_{\min}(L)$ estimated (and roughly extrapolated) from
Fig.~\ref{fig:condition_number_L}, and a \emph{high} value, set above it.
The accuracy is measured by the relative $\ell_2$-error between the
Clenshaw-based solution and the linear reference, 
evaluated as a function of intermediate time $t$.
We focus here on the most demanding case $T = 64$ ($N_t = 128$).


As expected, when $\kappa_{\mathrm{QSVT}}$ is set below the true effective
condition number, 
the error remains at the $10^{-1}$ level or above, regardless of $d_{\mathrm{QSVT}}$.
In contrast, when $\kappa_{\mathrm{QSVT}}$ exceeds the true value,
the error drops to the $10^{-1}$ level or below for sufficiently large $d_{\mathrm{QSVT}}$.
Moreover, the rate at which the error decreases with increasing $d_{\mathrm{QSVT}}$ is
substantially larger for the high-$\kappa_{\mathrm{QSVT}}$ cases, consistent with
the polynomial approximation theory~\eqref{eq:degree_polynomial}: when
$\kappa_{\mathrm{QSVT}}$ is underestimated, increasing $d_{\mathrm{QSVT}}$ alone cannot
compensate for the mismatch.

\begin{figure}[htbp]
  \centering
  \begin{subfigure}[b]{0.31\linewidth}
    \centering
    \includegraphics[width=\linewidth]{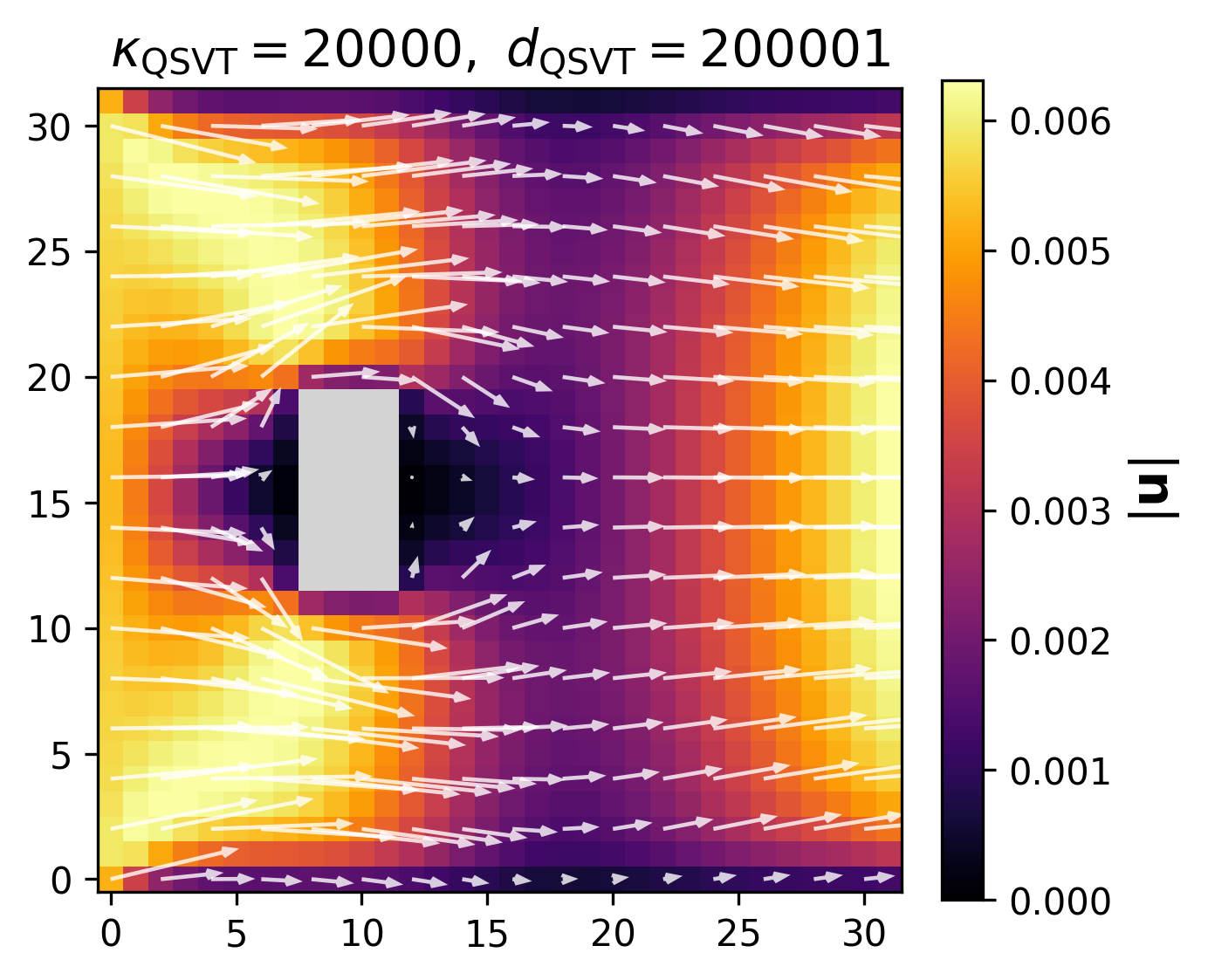}
    \subcaption{$\kappa_{\mathrm{QSVT}}$ not large enough 
    }
    \label{fig:velocity_under_kappa}
  \end{subfigure}
  \hfill
  \begin{subfigure}[b]{0.31\linewidth}
    \centering
    \includegraphics[width=\linewidth]{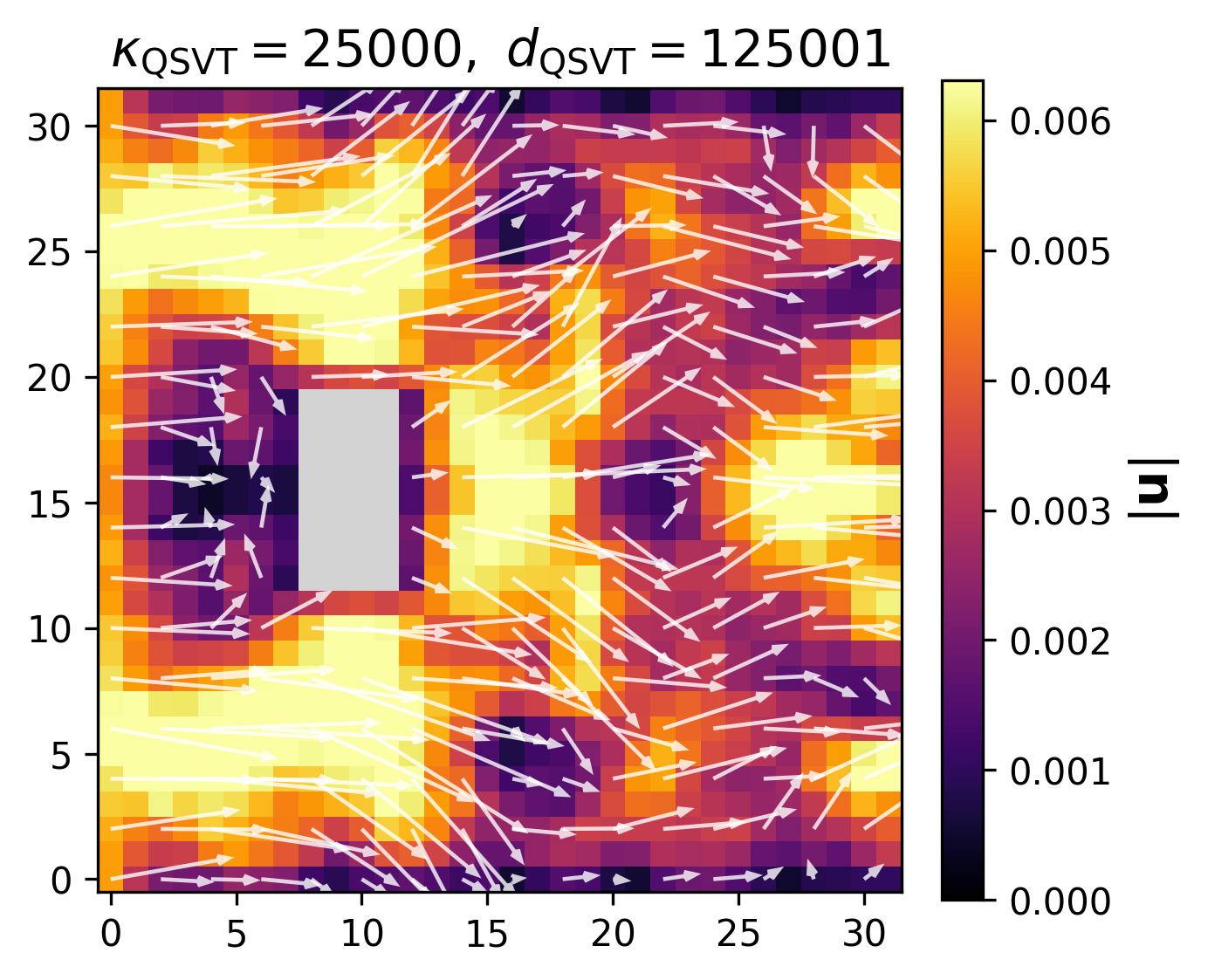}
    \subcaption{$d_{\mathrm{QSVT}}$ not large enough .
    }
    \label{fig:velocity_under_degree}
  \end{subfigure}
  \hfill
  \begin{subfigure}[b]{0.31\linewidth}
    \centering
    \includegraphics[width=\linewidth]{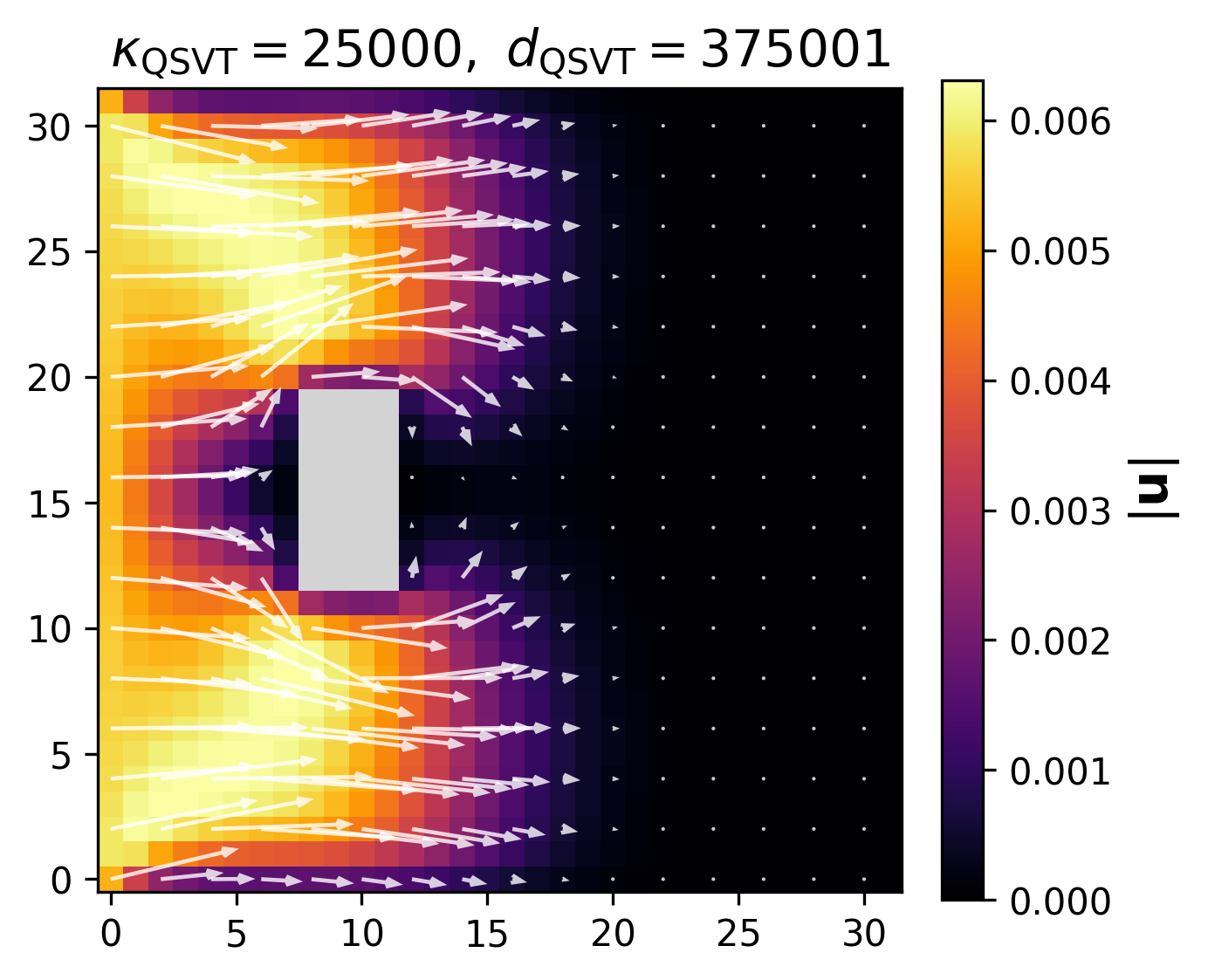}
    \subcaption{Both sufficiently large 
    }
    \label{fig:velocity_sufficient}
  \end{subfigure}
  \caption{Velocity fields at $t = 25$ ($n_t = 50$) for $N_x = N_y = 32$, $T = 64$.}
  \label{fig:velocity_comparison}
\end{figure}

To visualize these effects in physical space, Fig.~\ref{fig:velocity_comparison}
compares the velocity fields at intermediate time 
for three representative parameter choices. 
When $\kappa_{\mathrm{QSVT}}$ is not large enough
(Fig.~\ref{fig:velocity_under_kappa}),
the flow field is broadly correct in the interior
but develops spurious artifacts near the outflow (right) boundary.
When the polynomial degree is not large enough
(Fig.~\ref{fig:velocity_under_degree}),
the entire velocity field is corrupted by oscillatory noise.
Only when both $\kappa_{\mathrm{QSVT}}$ and $d_{\mathrm{QSVT}}$ are sufficiently large
(Fig.~\ref{fig:velocity_sufficient})
does the Clenshaw-based solution achieve good agreement with the linear reference.

\newpage

\subsection{Quantum computational costs of $U_L$}

\begin{figure}[htbp]
  \centering
	\begin{minipage}{0.48\linewidth}
		\includegraphics[width=0.9\textwidth]{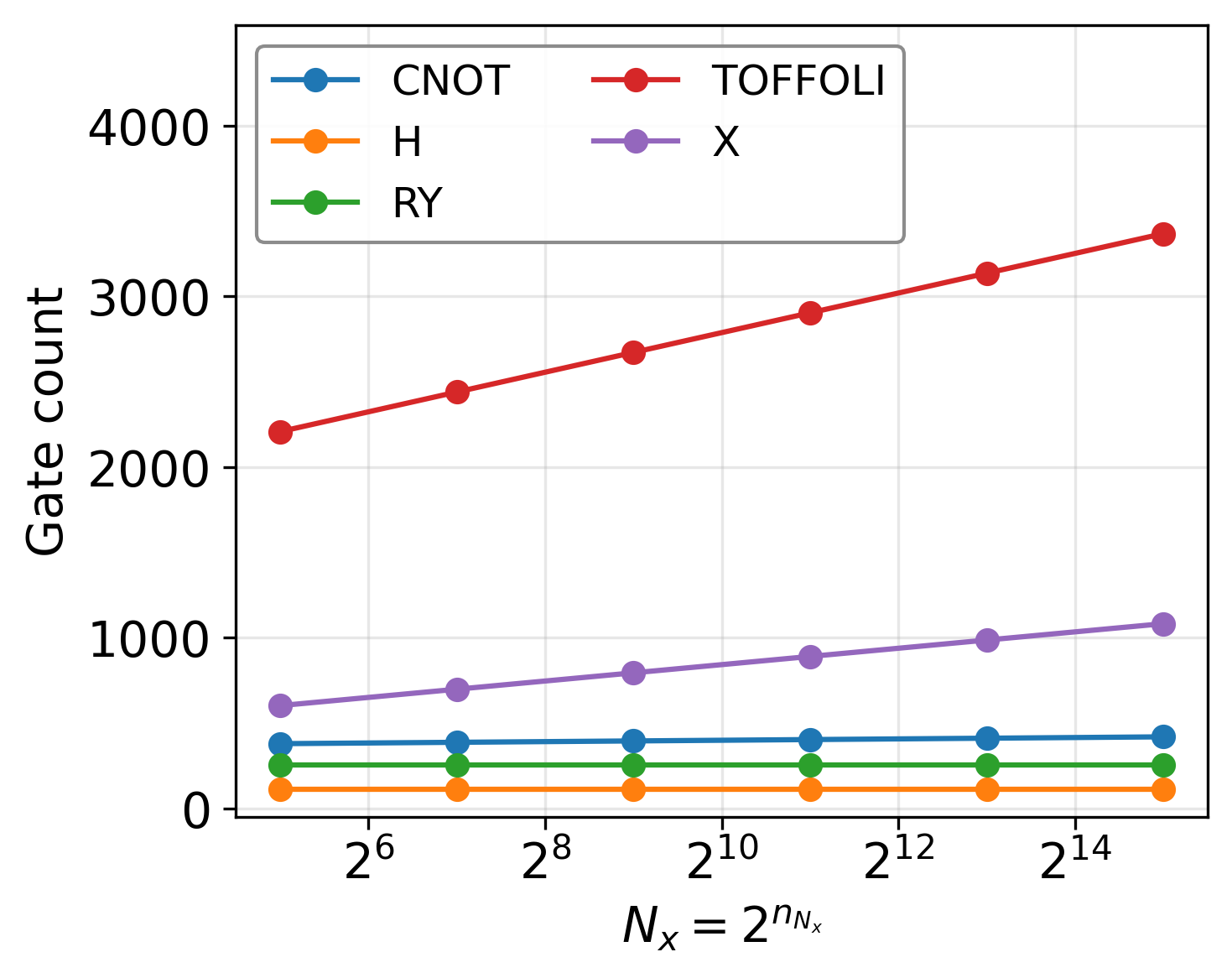}
		\subcaption{Gate counts vs.\,$N_x$ (for $N_t = 64$).}
		\label{fig:gate_costs_T64}
	\end{minipage}
	\begin{minipage}{0.48\linewidth}
		\includegraphics[width=0.9\textwidth]{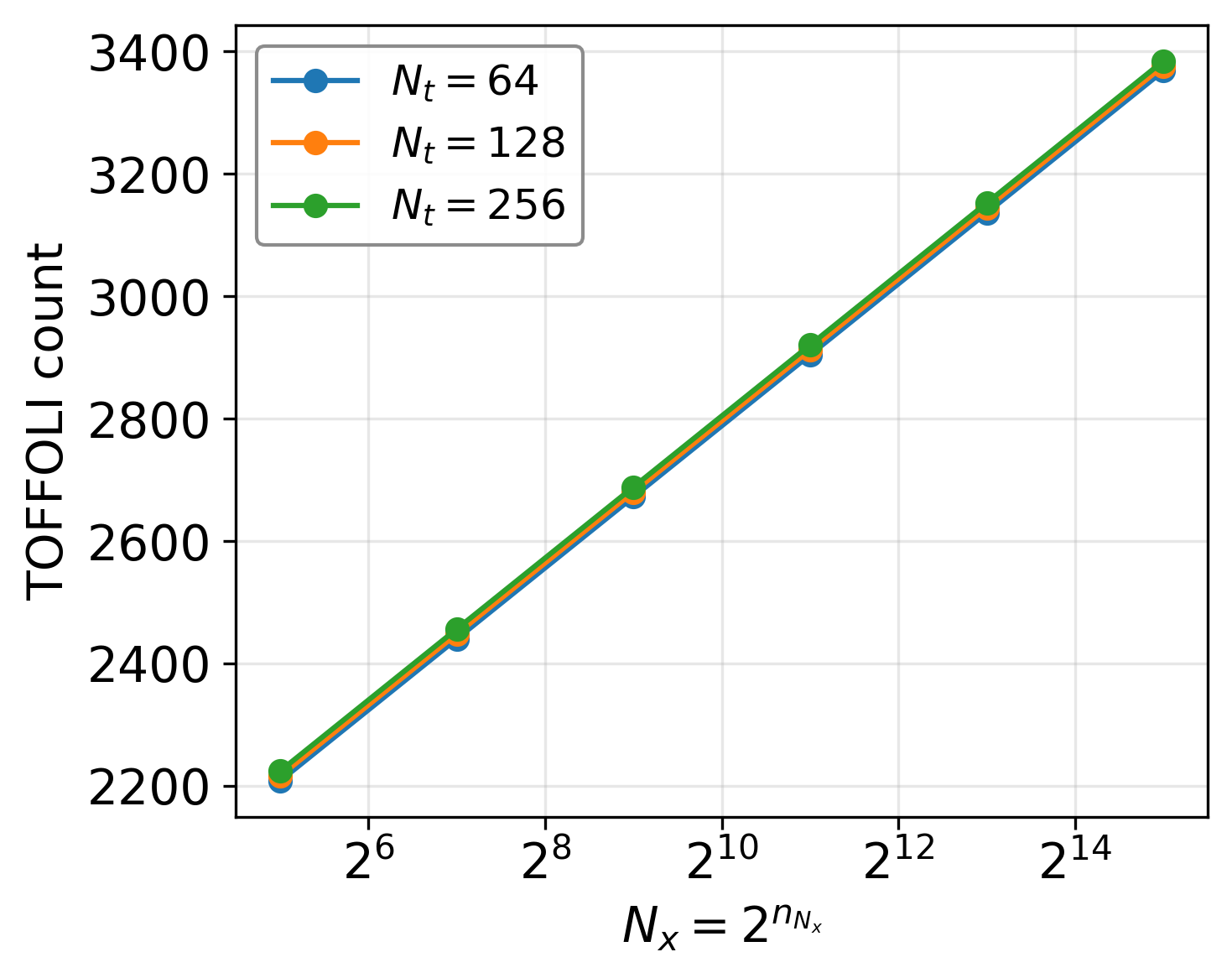}
		\subcaption{
			Toffoli count (vs.\,$N_x$) for different $N_t$'s.
		}
		\label{fig:toffoli_vs_T}
	\end{minipage}
  \caption{Quantum gate counts of $U_L$.
    }
  \label{fig:gate_counts}
\end{figure}

\begin{figure}[htbp]
  \centering
  \includegraphics[width=0.6\linewidth]{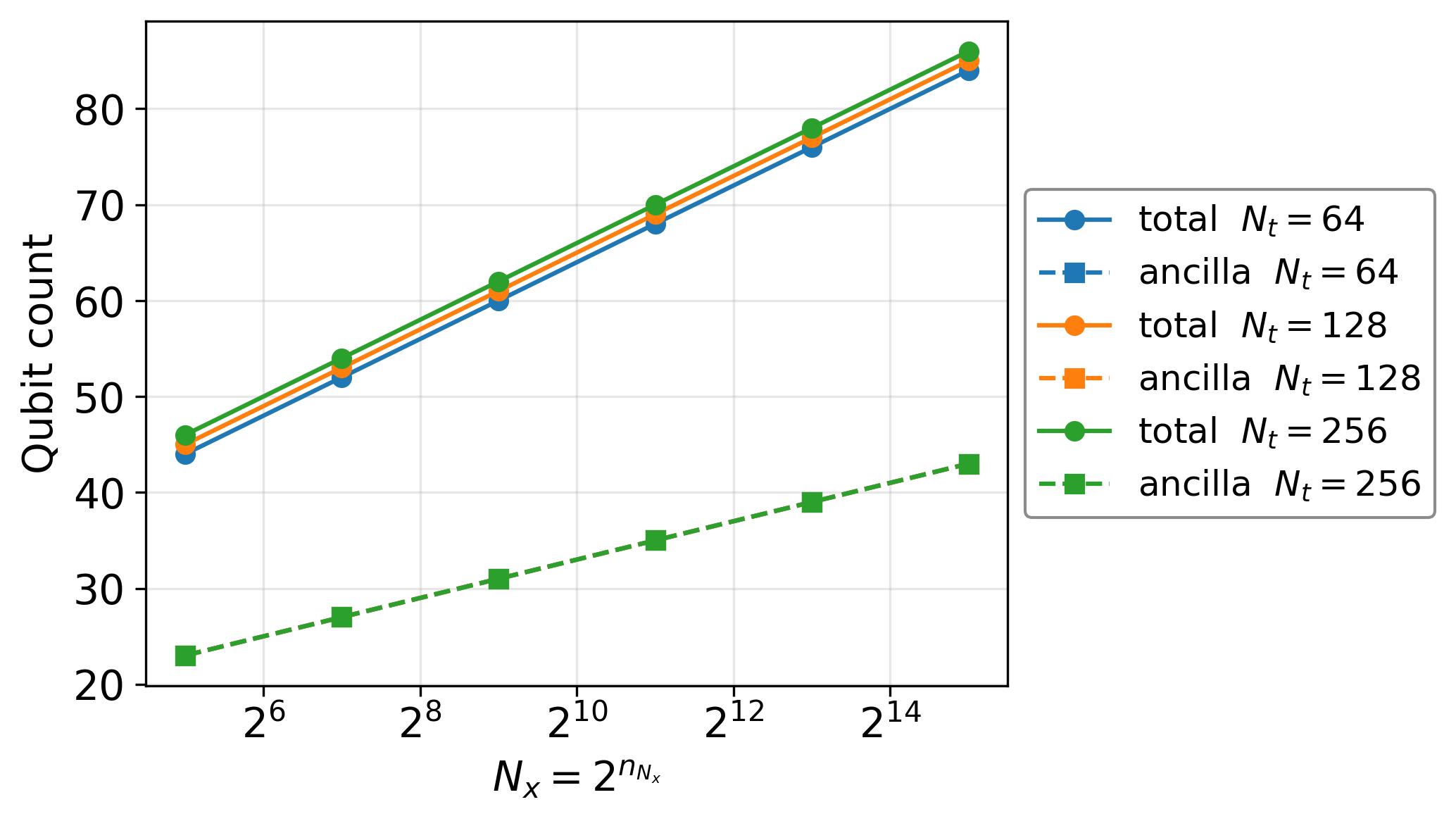}
  \caption{Qubit counts of $U_L$. 
  The ancilla qubit counts are identical for all $N_t$ values.} 
  \label{fig:qubit_counts}
\end{figure}

We analyze the quantum computational costs of the block-encoding $U_L$ by counting the required numbers of elementary gates and qubits for various $N_x\,(=N_y)$ and $N_t$. 
Figure~\ref{fig:gate_costs_T64} shows the count of each elementary gate for $N_t = 64$ as a function of $N_x = 2^{n_{N_x}}$.
The number of Toffoli gates grows logarithmically with $N_x$,  
while those of the other gates remain nearly constant. 
Figure~\ref{fig:toffoli_vs_T} further shows the Toffoli gate count for $N_t = 64, 128, 256$, 
which is almost independent of $N_t$.
This is consistent with the structure of $U_L$, in which the time-step dependence enters only through a small number of operations on ancilla qubits.
Figure~\ref{fig:qubit_counts} shows the total and ancilla qubit counts, both of which again grow logarithmically with $N_x$, in agreement with the asymptotic analysis of Sec.~\ref{sec:methods}.

\subsection{Fault-tolerant T-gate cost estimation}\label{subsec:Tgate_cost}

To translate the gate counts into fault-tolerant resource requirements,
we roughly estimate the total number of T~gates needed for a single application of the QSVT-based QLSA. 
Following \cite{Jennings2025-theory}, we decompose 
\begin{itemize}
	\item each Toffoli gate (in $U_L$) into 7~T~gates, and 
	\item each $R_Y$ gate (in $U_L$) and $R_z$ gate (in QSVT circuit) into $3\log_2(1/\epsilon_{\text{gate}})$~T~gates
	via the Clifford+T approximation~\cite{Ross2016-rotation}, where $\epsilon_{\text{gate}}$ is the per-gate synthesis precision.
\end{itemize}
Because the QSVT circuit calls $U_L$ a total of $d_{\mathrm{QSVT}}$~times, where $d_{\mathrm{QSVT}}$ is the
polynomial degree~\eqref{eq:degree_polynomial}, the total T-gate count is
\begin{equation}\label{eq:total_Tgate}
  N_{\mathrm{T}}^{(\text{total})}
  = N_{\mathrm{Toffoli}}^{(U_L)} \cdot d_{\mathrm{QSVT}} \cdot 7
  + \bigg(N_{R_Y}^{(U_L)} \cdot d_{\mathrm{QSVT}} + (d_{\mathrm{QSVT}} + 1) \bigg) \cdot 3\log_2\!\left(\frac{1}{\epsilon_{\text{gate}}}\right)
\end{equation}
where $N_{\mathrm{Toffoli}}^{(U_L)}$ (resp. $N_{R_Y}^{(U_L)}$) is the Toffoli (resp. $R_Y$) gate counts
of a single invocation of $U_L$ (Fig.~\ref{fig:gate_counts}). 
For the polynomial degree, we use $d_{\mathrm{QSVT}} = c \cdot \kappa_{\mathrm{QSVT}} + 1$,
dropping the logarithmic factor in Eq.~\eqref{eq:degree_polynomial} and absorbing
it into the proportionality constant~$c$.
The effective condition number is approximated as
$\kappa_{\mathrm{QSVT}} \approx 4 \cdot T^{1.2} \cdot 32$,
obtained by fitting $1/\sigma_{\min}(L)$ for the $N_x = 16$ obstacle
configuration in Fig.~\ref{fig:condition_number_L} and multiplying the
subnormalization factor $\alpha_L$ with an additional safety margin of~$1.2$.
Because this fit is based on a single small grid size and is extrapolated
to larger $T$ values, it should be regarded as an order-of-magnitude
approximation of $\kappa_{\mathrm{eff}}$.
The per-gate synthesis precision is set to
$\epsilon_{\text{gate}} = \epsilon_{\mathrm{base}} / d_{\mathrm{QSVT}}$
to keep the total synthesis error bounded by $\epsilon_{\mathrm{base}}$.
Based on the error analysis in Fig.~\ref{fig:clenshaw_error_vs_Nt},
we set $c = 10$, i.e.\ $d_{\mathrm{QSVT}} = 10\kappa_{\mathrm{QSVT}}+1$.
Under this choice, the QSVT approximation error is on the order of a few percent,
so we set $\epsilon_{\mathrm{base}} = 0.01$ to keep the gate-synthesis
error at a comparable level.

\begin{figure}[htbp]
  \centering
  \includegraphics[width=0.65\linewidth]{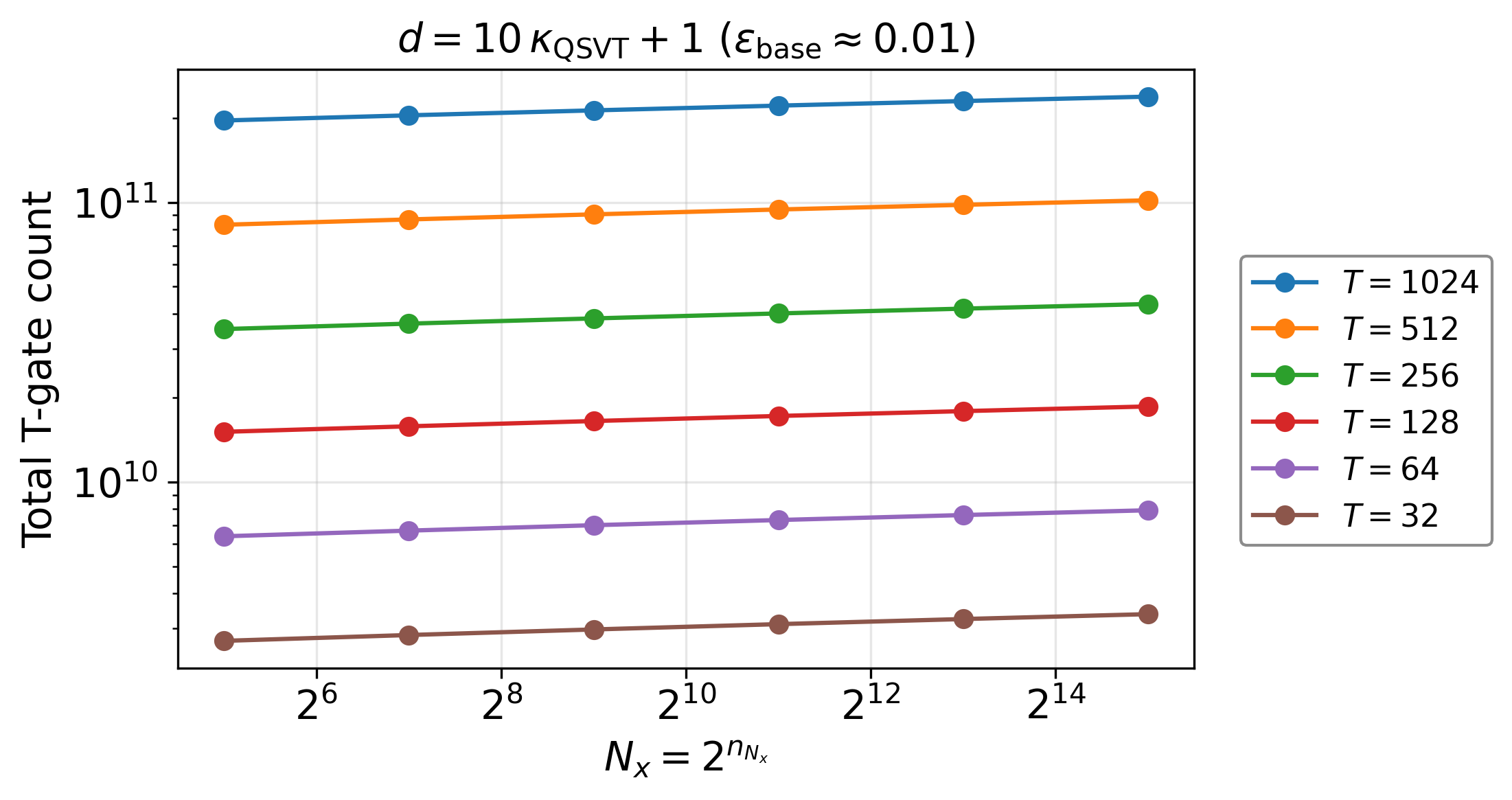}
  \caption{Estimated total T-gate count for a single QSVT-based linear system
    solve as a function of $N_x$, 
    with 
    $d_{\mathrm{QSVT}} = 10\,\kappa_{\mathrm{QSVT}}+1$.
    Under this setting, the QSVT approximation error and the total
    gate-synthesis error ($\epsilon_{\mathrm{base}} = 0.01$) are both
    on the order of a few percent.
    The T-gate count is dominated by the simulation time $T$ and depends only
    weakly on the spatial resolution $N_x$.}
  \label{fig:total_Tgate_cost}
\end{figure}

Figure \ref{fig:total_Tgate_cost} shows the resulting T-gate counts.
The total T-gate count ranges from approximately $10^{10}$ to $10^{12}$,
depending primarily on $T$: each doubling of $T$ increases the count by roughly
an order of magnitude.
In contrast, the dependence on $N_x$ is weak:
the curves are nearly flat, consistent with the logarithmic scaling of the gate cost
(Fig.~\ref{fig:gate_counts}) and the finding that
$\kappa_{\mathrm{eff}}$ is driven by $T$ rather than $N_x$
(Sec.~\ref{subsec:condition_number}).
We stress that these estimates are order-of-magnitude figures
that rely on several approximations, notably the extrapolated
$\kappa_{\mathrm{QSVT}}$ fit, the choice of the proportionality constant~$c$,
and the asymptotic gate-synthesis cost, and are intended to illustrate the
relative scaling with $T$ and $N_x$ rather than to provide precise resource
counts.
Moreover, they do not include the additional overhead from amplitude estimation
to be discussed in Sec.~\ref{sec:discussion}.

\section{Discussion}\label{sec:discussion}

\paragraph{Comparison with \cite{Jennings2025-theory,Jennings2025-apps}.}
As discussed in Sec.~\ref{sec:intro}, the present work extends the framework of \cite{Jennings2025-theory,Jennings2025-apps} in two directions: 
\begin{itemize}
\item[(i)] a characterization of how the effective condition number $\kappa_{\mathrm{QSVT}}$ (hence the number of queries to the block-encoding unitary) depends on $N_t$ and $N_x$ separately, and
\item[(ii)] an explicit gate-level construction of the block-encoding of the system matrix that incorporates realistic boundary conditions and obstacle geometry. 
\end{itemize}

Regarding~(i), \cite{Jennings2025-theory} concluded that quantum advantage at high Reynolds numbers is at most polynomial over classical CFD.
Our independent analysis (Sec.~\ref{subsec:condition_number}) provides a more detailed picture: $\kappa_{\mathrm{QSVT}}$ grows strongly with $N_t$ but depends only weakly on $N_x$.
At high Reynolds numbers, achieving a given physical simulation time requires resolving finer spatial and temporal scales, which translates to a larger $N_t$ per advection time and thus a polynomially growing $\kappa_{\mathrm{QSVT}}$.
Since the QSVT polynomial degree scales as $\widetilde{\mathcal{O}}(\kappa_{\mathrm{QSVT}})$, this drives up the overall quantum resource cost, consistent with their conclusion.
In contrast, when $N_t$ is fixed and the domain size (in effect $N_x$) is increased, the condition number remains nearly constant, so the quantum computational cost grows only mildly with system size.

Regarding~(ii), the gate-level construction in Sec.~\ref{subsec:quantum_circuit} shows that the quantum computational cost scales as $\mathcal{O}(\log N_x)$, so the marginal cost of increasing spatial resolution is low.
The fault-tolerant T-gate cost estimation in Sec.~\ref{subsec:Tgate_cost} quantitatively confirms this: the total T-gate count for a single QSVT-based solve 
is dominated by the simulation time $T$ rather than $N_x$.

Together, these two observations suggest that the most favorable regime for the proposed algorithm is one in which a large spatial domain is simulated over a moderate number of time steps.
Concrete examples include computing ensemble statistics by superposing runs with slightly different initial conditions, or exploring design variations, such as small perturbations to the obstacle geometry, within a single quantum computation.
Even under the constraint of a limited number of time steps, the logarithmic scaling with $N_x$ makes it possible to accommodate a large number of lattice points, or equivalently a large number of ensemble members, at modest additional quantum cost.

\paragraph{Algorithmic limitations and future improvements.}
The present implementation adopts the first-order Carleman truncation ($N_C = 1$), which linearizes the nonlinear LBE by retaining only the leading-order term.
While this is sufficient for the low-Ma and low-Re regime, 
it introduces a truncation error that grows with the Mach number.
Higher-order Carleman truncation ($N_C \geq 2$) can reduce this error at the cost of an exponential increase in the state-space dimension. 
The work of \cite{Jennings2025-theory} developed scalable block-encoding strategies and presented gate-cost estimates for arbitrary truncation orders $N_C$. Extending our circuit-level framework to support $N_C \geq 2$ is an important direction for future work.
We note that for the first-order truncation adopted in this work ($N_C = 1$), the shifted-LBM equilibrium distribution introduced in \cite{Jennings2025-theory} is not required.
However, at higher truncation orders, the shifted formulation becomes essential: it keeps the norm of the state vector bounded as $\mathcal{O}(1)$, thereby controlling the overhead of extracting classical information from the quantum state.

\paragraph{Applicability at high Reynolds numbers.}
A fundamental challenge for Carleman-linearization-based quantum algorithms is that the convergence of the Carleman series is not guaranteed at high Reynolds numbers.
In the LBM setting, this translates to a restriction on the Mach number of the inflow, which must remain sufficiently small for the Carleman expansion to converge.
The present study is conducted at $\mathrm{Re} = 1$ and $\mathrm{Ma} = 0.01$, well within this regime.
Extending the algorithm to turbulent, high-Reynolds-number flows requires either higher-order Carleman truncation with provable convergence bounds or fundamentally different linearization strategies, and is left for future work.

\paragraph{End-to-end resource estimates.}
The fault-tolerant T-gate cost estimation in Sec.~\ref{subsec:Tgate_cost} provides a concrete resource estimate for the block-encoding-plus-QSVT portion of the algorithm. 
However, a complete end-to-end resource estimate must also account for several additional costs that are not addressed in this work.
First, extracting flow observables such as the drag force from the quantum state requires quantum amplitude estimation (as described, for example, in \cite{Penuel2024-cl}), 
which introduces an additional multiplicative overhead of $\mathcal{O}(\kappa_{\mathrm{QSVT}} / \epsilon)$ relative to a single application of the QLSA, where $\epsilon$ is the target precision.
Combined with the QSVT polynomial degree $d_{\mathrm{QSVT}} = \widetilde{\mathcal{O}}(\kappa_{\mathrm{QSVT}})$, the total number of queries to the block-encoding 
scales as $\widetilde{\mathcal{O}}(\kappa_{\mathrm{QSVT}}^2 / \epsilon)$, which grows quadratically with $N_t$ up to logarithmic factors.
Second, the cost of preparing the initial quantum state encoding the initial condition $\bm{b}_L$ has not been included in our estimates. This cost depends on the structure of the initial data and may itself require nontrivial quantum circuits.
For uniform initial conditions such as the zero-velocity state adopted in this work, however, state preparation is straightforward, as the same value is assigned to all lattice points and the preparation cost does not scale with $N_x$.
Accounting for these readout and state-preparation costs would substantially increase the total quantum resource requirements beyond the estimates in Sec.~\ref{subsec:Tgate_cost} and should be incorporated in future end-to-end analyses.

\section{Conclusion}\label{sec:conclusion}
In this work, we developed and validated a quantum algorithm for simulating two-dimensional flow around an obstacle based on Carleman-linearized LBM and QSVT-based matrix inversion.
The central contribution is an explicit gate-level construction of the quantum circuit for the system matrix block-encoding $U_L$, which faithfully encodes the full time-discretized LBM dynamics---including inflow, outflow, and no-slip boundary conditions as well as an obstacle geometry---without relying on high-level oracle abstractions.
To our knowledge, this is the first demonstration of a QLBM circuit that incorporates realistic inflow and outflow boundary conditions at the gate level.

Furthermore, a systematic numerical study of the effective condition number $\kappa_{\mathrm{QSVT}}$ of the system matrix $L$ reveals that $\kappa_{\mathrm{QSVT}}$ grows strongly with the number of time steps $N_t$, whereas its dependence on the number of lattice points $N_x$ is comparatively weak.
This finding identifies $N_t$ as the dominant driver of quantum resource costs in the global-in-time formulation, and suggests that the proposed algorithm is most competitive in the regime of large spatial resolution but moderate temporal extent.

We validated the quantum algorithm through state-vector simulation and also confirmed that the Clenshaw algorithm faithfully emulates the QSVT polynomial transformation.
The latter classical emulation allowed us to study larger problem sizes than are accessible to full state-vector simulation, and to directly observe how the choice of condition number parameter 
$\kappa_{\mathrm{QSVT}}$ and polynomial degree $d_{\mathrm{QSVT}}$ affects the accuracy of the reconstructed flow field.
This provides a practical tool for calibrating quantum resource parameters before running the algorithm on actual quantum hardware.

Finally, a gate-count analysis of the block-encoding $U_L$ shows that the total number of gates scales approximately logarithmically with $N_x$, in agreement with the asymptotic analysis.
When translated into fault-tolerant T-gate costs, the total count ranges from $10^{10}$ to $10^{12}$ for the problem sizes considered ($N_x = N_y \leq 10^{16}$ and $T \lesssim 10^{3}$), with the simulation time $T$ remaining the dominant cost driver rather than the spatial resolution $N_x$.
This logarithmic scaling is encouraging for large-scale applications, as it implies that substantially increasing the spatial resolution of the simulation requires only a modest increase in quantum circuit complexity.

\section*{Acknowledgments}
The authors thank Hayato Higuchi and Kouki Nakamura for helpful discussions.
This work was performed using the JAXA Supercomputer System `JSS3'. This work was also supported by MEXT as ``Program for Promoting Researches on the Supercomputer Fugaku'' and used computational resources of supercomputer Fugaku provided by the RIKEN Center for Computational Science (Project ID: ra250010). The authors would like to thank Masafumi Yamazaki for his valuable support on the use of Fugaku supercomputer system.

\newpage

\appendix

\section{Notation}\label{appendix:notations}

\paragraph{Lattice Boltzmann method}
	\begin{longtable}{ccl}
		$D$ & : & spatial dimension\\
		$Q$ & : & number of discrete velocities\\
		$N_x, N_y$ & : & number of lattice nodes in each direction\\
		$N$ & : & total number of lattice nodes ($N = N_x \times N_y$)\\
		$\bm{x}_n$ & : & position of the $n$-th lattice node\\
		$\bm{c}_q$ & : & discrete velocity vector for index $q$\\
		$w_q$ & : & weight associated with $\bm{c}_q$\\
		$f_q(\bm{x}_n, t)$ & : & discrete distribution function\\
		$f_q^{\text{eq.}}(\bm{x}_n, t)$ & : & equilibrium distribution function\\
		$f_q^*(\bm{x}_n, t)$ & : & post-collision distribution function\\
		$\rho(\bm{x}_n, t)$ & : & macroscopic density\\
		$\bm{u}(\bm{x}_n, t)$ & : & macroscopic velocity\\
		$c_{\text{s}}$ & : & lattice speed of sound ($= 1/\sqrt{3}$ in lattice units)\\
		$\nu$ & : & kinematic viscosity\\
		$\tau$ & : & relaxation time\\
		$\Omega_q$ & : & collision operator\\
		$\bar{q}$ & : & index of the velocity opposite to $\bm{c}_q$\\
		$\bm{u}_{\text{in}}$ & : & prescribed inflow velocity\\
		$\rho_{\text{in}}$ & : & density at the boundary node
	\end{longtable}

\paragraph{Matrix-vector formulation and Carleman linearization}
	\begin{longtable}{ccl}
		$\bm{f} \in \mathbb{R}^{NQ}$ & : & vectorized distribution function\\
		$F_1, F_2, F_3$ & : & collision matrices (linear, quadratic, cubic)\\
		$S$ & : & streaming permutation matrix\\
		$\bm{b} \in \mathbb{R}^{NQ}$ & : & inflow forcing vector\\
		$A^{(1)} = S(I + F_1)$ & : & collision-streaming matrix (first-order)\\
		$q, n$ & : & pre-collision velocity and node indices (input)\\
		$q^{*}$ & : & post-collision velocity index (node unchanged by collision)\\
		$q_{\mathrm{out}}, n_{\mathrm{out}}$ & : & post-streaming velocity and node indices (output)\\
		$C_{q^{*}q}$ & : & collision matrix element of $I + F_1$\\
		$N_C$ & : & Carleman truncation order\\
		$\bm{y}(t)$ & : & Carleman vector (stacked tensor powers of $\bm{f}$)\\
		$d_C$ & : & dimension of the Carleman vector\\
		$\mathcal{C}, \mathcal{S}$ & : & Carleman collision and streaming matrices\\
		$\bm{b}_C$ & : & Carleman-extended forcing vector\\
		$h$ & : & step-size parameter for interpolated update\\
		$\widetilde{A}$ & : & interpolated evolution matrix, $(1-h)I + hA^{(1)}$
	\end{longtable}

\paragraph{Linear system}
	\begin{longtable}{ccl}
		$N_t$ & : & number of time steps\\
		$T$ & : & total simulation time ($= N_t \cdot h$)\\
		$W$ & : & idling phase parameter\\
		$L$ & : & block bidiagonal matrix of the linear system\\
		$\bm{y}$ & : & solution vector\\
		$\bm{b}_L$ & : & right-hand side vector
	\end{longtable}

\paragraph{Quantum circuit}
	\begin{longtable}{ccl}
		$n_{N_x}, n_{N_y}$ & : & number of qubits for spatial registers\\
		$n_Q$ & : & number of qubits for velocity register\\
		$\widetilde{Q}$ & : & padded number of velocities ($= 2^{n_Q}$)\\
		$n_{N_t}$ & : & number of qubits for time-step register\\
		$\mathrm{BC}$ & : & boundary-condition type\\
		$\alpha_A$ & : & subnormalization factor of $A^{(1)}$ block-encoding\\
		$\alpha_L$ & : & subnormalization factor of $L$ block-encoding\\
		$c_L$ & : & normalization factor for individual $R_Y$ rotation amplitudes in $U_L$\\
		$\kappa_{\mathrm{QSVT}}$ & : & condition number parameter for QSVT polynomial\\
		$d_{\mathrm{QSVT}}$ & : & polynomial degree parameter for QSVT\\
		$U_{A^{(1)}}$ & : & block-encoding unitary for $A^{(1)}$\\
		$U_L$ & : & block-encoding unitary for $L$
	\end{longtable}

\newpage

\renewcommand{\baselinestretch}{1.1}
\bibliographystyle{ytamsalpha}
\bibliography{bib}

\newcommand{\etalchar}[1]{$^{#1}$}
\providecommand{\bysame}{\leavevmode\hbox to3em{\hrulefill}\thinspace}
\providecommand{\MR}{\relax\ifhmode\unskip\space\fi MR }
\providecommand{\MRhref}[2]{%
  \href{http://www.ams.org/mathscinet-getitem?mr=#1}{#2}
}
\providecommand{\href}[2]{#2}
\providecommand{\doihref}[2]{\href{#1}{#2}}
\providecommand{\arxivfont}{\tt}
\begin{thebibliography}{JKL{\etalchar{+}}25b}

\bibitem[Ber10]{Berry2014-zu}
D.~W. Berry, \emph{High-order quantum algorithm for solving linear differential equations}, \doihref{https://doi.org/10.1088/1751-8113/47/10/105301}{J. Phys. A Math. Theor. \textbf{47} (2014) 105301}, \href{https://arxiv.org/abs/1010.2745}{{\arxivfont arXiv:1010.2745 [quant-ph]}}.

\bibitem[BGK54]{Bhatnagar1954-bgk}
P.~L. Bhatnagar, E.~P. Gross, and M.~Krook, \emph{{A Model for Collision Processes in Gases. I. Small Amplitude Processes in Charged and Neutral One-Component Systems}}, \doihref{https://doi.org/10.1103/PhysRev.94.511}{Phys. Rev. \textbf{94} (1954) 511--525}.

\bibitem[BS24]{BerntsonSünderhauf:2024}
B.~K. Berntson and C.~Sünderhauf, \emph{Complementary polynomials in quantum signal processing}, \doihref{https://doi.org/10.1007/s00220-025-05302-9}{Commun. Math. Phys. \textbf{406} (2025) }, \href{https://arxiv.org/abs/2406.04246}{{\arxivfont arXiv:2406.04246 [quant-ph]}}.

\bibitem[Bud21]{Budinski2021-iy}
L.~Budinski, \emph{Quantum algorithm for the advection-diffusion equation simulated with the lattice {B}oltzmann method}, \doihref{https://doi.org/10.1007/s11128-021-02996-3}{Quantum Inf. Process. \textbf{20} (2021) 1--17}.

\bibitem[Car32]{Carleman1932-zr}
T.~Carleman, \emph{Application de la théorie des équations intégrales linéaires aux systèmes d'équations différentielles non linéaires}, \doihref{https://doi.org/10.1007/bf02546499}{Acta Math. \textbf{59} (1932) 63--87}.

\bibitem[Cle55]{Clenshaw1955-iy}
C.~W. Clenshaw, \emph{A note on the summation of {Chebyshev} series}, \doihref{https://doi.org/10.1090/S0025-5718-1955-0071856-0}{Math. Comp. \textbf{9} (1955) 118--120}.

\bibitem[GKS21]{GriblingKerenidisSzilágyi}
S.~Gribling, I.~Kerenidis, and D.~Szil\'{a}gyi, \emph{An optimal linear-combination-of-unitaries-based quantum linear system solver}, \doihref{https://doi.org/10.1145/3649320}{ACM Trans. Quantum Comput. \textbf{5} (2024) 1--23}, \href{https://arxiv.org/abs/2109.04248}{{\arxivfont arXiv:2109.04248 [quant-ph]}}.

\bibitem[GSLW18]{GilyenSuLowWiebe:2018}
A.~Gily\'{e}n, Y.~Su, G.~H. Low, and N.~Wiebe, \emph{{Q}uantum {S}ingular {V}alue {T}ransformation and beyond: Exponential improvements for quantum matrix arithmetics}, \doihref{https://doi.org/10.1145/3313276.3316366}{STOC 2019: Proc. 51st Ann. ACM SIGACT Symp. on Theory of Computing (2019) 193--204}, \href{https://arxiv.org/abs/1806.01838}{{\arxivfont arXiv:1806.01838 [quant-ph]}}.

\bibitem[JKL{\etalchar{+}}25a]{Jennings2025-theory}
D.~Jennings, K.~Korzekwa, M.~Lostaglio, R.~Ashworth, E.~Marsili, and S.~Rolston, \emph{An end-to-end quantum algorithm for nonlinear fluid dynamics with bounded quantum advantage}, \href{https://arxiv.org/abs/2512.03758}{{\arxivfont arXiv:2512.03758 [quant-ph]}}.

\bibitem[JKL{\etalchar{+}}25b]{Jennings2025-apps}
D.~Jennings, K.~Korzekwa, M.~Lostaglio, P.~Mannix, R.~Ashworth, E.~Marsili, and S.~Rolston, \emph{Simulating non-trivial incompressible flows with a quantum lattice {B}oltzmann algorithm}, \href{https://arxiv.org/abs/2512.05781}{{\arxivfont arXiv:2512.05781 [physics.flu-dyn]}}.

\bibitem[KKK{\etalchar{+}}16]{Kruger2016-lj}
T.~Kr{\"u}ger, H.~Kusumaatmaja, A.~Kuzmin, O.~Shardt, G.~Silva, and E.~M. Viggen, \doihref{https://doi.org/10.1007/978-3-319-44649-3}{\emph{{The Lattice Boltzmann Method: Principles and Practice}}}, {Graduate Texts in Physics}, Springer International Publishing, 2016.

\bibitem[LYW{\etalchar{+}}23]{Li2025-ox}
X.~Li, X.~Yin, N.~Wiebe, J.~Chun, G.~K. Schenter, M.~S. Cheung, and J.~Mülmenstädt, \emph{Potential quantum advantage for simulation of fluid dynamics}, \doihref{https://doi.org/10.1103/physrevresearch.7.013036}{Phys. Rev. Res. \textbf{7} (2025) 013036}, \href{https://arxiv.org/abs/2303.16550}{{\arxivfont arXiv:2303.16550 [quant-ph]}}.

\bibitem[MPS{\etalchar{+}}24]{morales2025quantumlinearsolverssurvey}
M.~E.~S. Morales, L.~Pira, P.~Schleich, K.~Koor, P.~C.~S. Costa, D.~An, A.~Aspuru-Guzik, L.~Lin, P.~Rebentrost, and D.~W. Berry, \emph{Quantum linear system solvers: A survey of algorithms and applications}. \href{https://arxiv.org/abs/2411.02522}{{\arxivfont arXiv:2411.02522 [quant-ph]}}.

\bibitem[MSL{\etalchar{+}}15]{Mezzacapo2015-hw}
A.~Mezzacapo, M.~Sanz, L.~Lamata, I.~L. Egusquiza, S.~Succi, and E.~Solano, \emph{Quantum simulator for transport phenomena in fluid flows}, \doihref{https://doi.org/10.1038/srep13153}{Sci. Rep. \textbf{5} (2015) 13153}, \href{https://arxiv.org/abs/1502.00515}{{\arxivfont arXiv:1502.00515 [quant-ph]}}.

\bibitem[PKJ{\etalchar{+}}24]{Penuel2024-cl}
J.~Penuel, A.~Katabarwa, P.~D. Johnson, P.~Kuklinski, B.~Rempfer, C.~Farquhar, Y.~Cao, and M.~C. Garrett, \emph{Detailed assessment of calculating drag force with quantum computers: Explicit time-evolution precludes exponential advantage for nonlinear differential equations}, \href{https://arxiv.org/abs/2406.06323}{{\arxivfont arXiv:2406.06323 [quant-ph]}}.

\bibitem[RS14]{Ross2016-rotation}
N.~J. Ross and P.~Selinger, \emph{Optimal ancilla-free {Clifford+T} approximation of z-rotations}, \doihref{https://doi.org/10.26421/QIC16.11-12-1}{Quantum Inf. Comput. \textbf{16} (2016) 901--953}, \href{https://arxiv.org/abs/1403.2975}{{\arxivfont arXiv:1403.2975 [quant-ph]}}.

\bibitem[SS23]{Sanavio2024-je}
C.~Sanavio and S.~Succi, \emph{Lattice {Boltzmann-Carleman} quantum algorithm and circuit for fluid flows at moderate {R}eynolds number}, \doihref{https://doi.org/10.1116/5.0195549}{AVS Quantum Sci. \textbf{6} (2024) 023802}, \href{https://arxiv.org/abs/2310.17973}{{\arxivfont arXiv:2310.17973 [quant-ph]}}.

\bibitem[SSR{\etalchar{+}}25]{Sanavio2025-gn}
C.~Sanavio, W.~A. Simon, A.~Ralli, P.~Love, and S.~Succi, \emph{Carleman-lattice-{B}oltzmann quantum circuit with matrix access oracles}, \doihref{https://doi.org/10.1063/5.0254588}{Phys. Fluids \textbf{37} (2025) 037123}, \href{https://arxiv.org/abs/2501.02582}{{\arxivfont arXiv:2501.02582 [quant-ph]}}.

\bibitem[Suc18]{Succi2018-yi}
S.~Succi, \doihref{https://doi.org/10.1093/oso/9780199592357.001.0001}{\emph{{The Lattice Boltzmann Equation: For Complex States of Flowing Matter}}}, Oxford University Press, 2018.

\end{thebibliography}

\end{document}